\documentstyle[twocolumn,prb,aps,epsf]{revtex}

\def\grad{\mathop{\mathrm{grad}}\nolimits}

\def\dfrac#1#2{\displaystyle{\frac{\displaystyle{#1}}{\displaystyle{#2}}}}

\def\pdfrac#1#2{\dfrac{\partial#1}{\partial#2}}

\def\mbf#1{\mbox{\boldmath $#1$}}
\newcommand{\circlenum}[1]{{\ooalign{%
           \hfill$\scriptstyle#1$\hfill\crcr$\bigcirc$}}}

\begin{document}
\draft
\title{Reconnection and acoustic emission of quantized vortices in
superfluid \\ by the numerical analysis of the Gross-Pitaevskii equation}
\author{Shin-ichiro Ogawa$^1$, Makoto Tsubota$^1$, Yuji Hattori$^2$}
\address{$^1$Department of Physics, Osaka City University,
Sumiyoshi-Ku, Osaka 588-8585, Japan \\
$^2$Department of Computer Aided Science, Kyusyu Institute of
Technology \\
Sensui 1-1, Tobata-ku, Kitakyushu 804-8550, Japan}
\date{Received
\hspace{25mm}
}
\wideabs{
\maketitle
%%%%%%%%%%%%%%%%%%%% Abstract %%%%%%%%%%%%%%%%%%%%%%
\begin{abstract}
 We study numerically the reconnection of quantized vortices and the
 concurrent acoustic emission by the analysis of the Gross-Pitaevskii
 equation. Two quantized vortices reconnect following the process
 similar to classical vortices; they approach, twist themselves locally
 so that they become anti-parallel at the closest place, reconnect and
 leave separately.The investigation of the motion of the singular lines
 where the amplitude of the wave function vanishes in the vortex cores
 confirms that they follow the above scenario by reconnecting at a
 point. This reconnection is not contradictory to the Kelvin's circulation
 theorem, because the potential of the superflow field becomes undefined
 at the reconnection point. When the locally anti-parallel part of the
 vortices becomes closer than the healing length, it moves with the
 velocity comparable to the sound velocity, emits the sound waves and
 leads to the pair annihilation or reconnection; this phenomena is
 concerned with the Cherenkov resonance. The vortices are broken up to
 smaller vortex loops through a series of reconnection, eventually
 disappearing with the acoustic emission. This may correspond to the
 final stage of the vortex cascade process proposed by Feynman. The
 change in energy components, such as the quantum, the compressible and
 incompressible kinetic energy is analyzed for each dynamics. The
 propagation of the sound waves not only appears in the profile of the
 amplitude of the wave function but also affects the field of its phase,
 transforming the quantum energy due to the vortex cores to the kinetic
 energy of the phase field.
\end{abstract}
}

\section{Introduction}
The numerical simulations of the quantized vortex dynamics in superfluid
are classified into two methods\cite{don}.
The first is the vortex filament formulation and the second
is the numerical analysis of the Gross-Pitaevskii equation(GPE).
The former simulation pioneered by Schwarz \cite{Schwarz} is useful for
superfluid $^4$He,
since the core of quantized vortex is very thin, {\it i.e.} of the order
of atomic size.
In this formulation, the motion of vortices is caused by the local
and the nonlocal induced velocity fields subject to the Biot-Savart law.
When two vortices approach each other within a critical distance,
they are assumed to reconnect.
The results of this formulation agree excellently with
the experimental results of the vortex tangle in superfluid turbulence.
However, the 
significant phenomena concerned with the core structure, such as vortex
nucleation, annihilation and reconnection, cannot be described by this
formulation.

On the other hand, the analysis of the GPE, which is discussed in this
paper, can investigate and analyze the superfluid dynamics including
the motion of vortex cores.
Koplik and Levine solved
numerically the GPE and showed two closed vortices
actually reconnected even in an inviscid fluid \cite{Koplik}.
%%%%%%%%%%%%%%%%%%%%%%%%%%%%%%%%%%%%%%%%%%%%%%%%%
However, they did not show the detail of the reconnection process
%%%%%%%%%%%%%%%%%%%%%%%%%%%%%%%%%%%%%%%%%%%%%%%%%
and the concurrent acoustic emission, {\it i.e.} the emission of
the condensate density waves.
This is a significant phenomenon in the vortex dynamics.

As well-known, the reconnection of vortices and the concurrent acoustic
emission occur in a classical viscous fluid\cite{clas}.
In this case, the viscosity plays an important role in the reconnection.
The deduction from this phenomena is not applicable to the inviscid
superfluid.
%%%%%%%%%%%%%%%%%%%%%%%%%%%%%%%%%%%
The important difference between classical and quantized vortex cores is
the density profile.
The quantized vortex has the core where the superfluid density vanishes.
On the other hand, the vortex core in an ordinary classical fluid
has the finite fluid density.
It is the core structure that causes the reconnection of quantized
vortices even in an inviscid fluid.

The reconnection and the concurrent acoustic emission
in superfluid are significantly related with the ``eddy viscosity''
which is a long-standing problem in superfluid physics.
The superfluid turbulent state\cite{tough} in a capillary flow
induces excess temperature and pressure differences between both
ends of the capillary. 
It is widely recognized that the temperature difference arises from
the mutual friction due to the scattering of the excitations by
the quantized vortex.
On the other hand, the pressure difference is described
phenomenologically by the eddy viscosity.
The eddy viscosity works for superfluid and reduces its total momentum,
but its origin has not been necessarily revealed.
The reconnection and the concurrent acoustic emission
reduce the energy of the quantized vortex, and then increase
the kinetic energy of the superfluid including the sound waves.
If the kinetic energy decays,
the total energy and momentum of the superfluid are reduced.
Therefore,
the reconnection and the concurrent acoustic emission may be
the origin of the eddy viscosity.

Recently Leadbeater {\it et al.}\cite{adams} found that a sound wave is
emitted when two vortex rings reconnect.
They calculate the energy of the sound wave by estimating
the reduction of the vortex line length.
Our works investigate the change of energy components when two vortices
reconnect, following the method of Nore {\it et al}\cite{Nore}.

In this paper, we show the detailed process of the reconnection of quantized
vortices by the numerical analysis of the GPE, and the concurrent acoustic
emission.
Section \ref{sba} describes
the basic equations and the numerical method.
In Sec. \ref{sound}, we show the acoustic emission when two
anti-parallel vortices approach and disappear.
This acoustic emission could be related with the Cherenkov
resonance\cite{Cherenkov}. The obtained solutions are concerned with
the stationary solitary waves studied by Jones and Roberts\cite{Jones}.
In Sec. \ref{rec1}, to show the detailed process of the
reconnection, we analyze the motion of two vortices which are initially
placed at a right angle.
Section \ref{rec2} describes the sequence of reconnections and
concurrent acoustic emissions. This process resembles
Feynman's cascade process\cite{Feynman}.
%%%%%%%%%%%%%%%%%%%%%%%%%%%%%%%%%%%%%%%%%%%%%%%%
Section \ref{energy} describes the change in energy components.
Considering the quantized vortex reconnection,
we need to discuss whether it contradicts the Kelvin's circulation theorem
that states the conservation law of the circulation in an ideal fluid,
which will be described in Sec. \ref{kel}.
%%%%%%%%%%%%%%%%%%%%%%%%%%%%%%%%%%%%%%%%%%%%%%%%
Section \ref{conc} is devoted to conclusions and discussions.

\section{Basic Equations and Numerical Method}\label{sba}
The GPE is
\begin{equation}
 i\hbar\pdfrac{\Psi(\mbf{x})}{t}=
  -\dfrac{\hbar^2}{2m}\nabla^2\Psi(\mbf{x})+g|\Psi(\mbf{x})|^2
  \Psi(\mbf{x}),
  \label{e07}
\end{equation}
where $\Psi(\mbf{x})$ is the macroscopic wave function,
and the chemical potential $\mu$ is renormalized by the global gauge
transformation 
$\Psi(\mbf{x})=\Psi'(\mbf{x})\exp\left(-i\mu t/\hbar\right)$.
The coefficient $g$ of the interaction is related with the $s$-wave
scattering length $a$ as $g=4\pi\hbar^2a/m$.
The GPE conserves the number of particles and the total energy given
by
\begin{equation}
 N = \int d^3x |\Psi(\mbf{x})|^2,
  \label{e08}
\end{equation}
\begin{equation}
 E_{\rm tot} = \int d^3x
  \left(\dfrac{\hbar^2}{2m}\left|\nabla\Psi(\mbf{x})\right|^2
   +g\left|\Psi(\mbf{x})\right|^4\right),
  \label{e08e}
\end{equation}
respectively.
This wave function is expressed in terms of a density and a phase by
Madelung's transformation
\begin{equation}
 \Psi(\mbf{x})=\sqrt{\rho(\mbf{x})}\exp\left(i\theta(\mbf{x})\right).
  \label{e09}
\end{equation}
The superfluid velocity is given by
$\mbf{v}(\mbf{x})=\hbar\nabla\theta/m$, where $m$ is the mass of bose
particles. 

The total energy $E_{\rm tot}$ can be decomposed as the following\cite{Nore}:
\begin{eqnarray}
 E_{\rm tot}&&=E_{\rm kin}+E_{\rm int}+E_{\rm q},
  \label{e10} \\
 E_{\rm kin}&&=\int d^3 x
  \dfrac{\hbar^2}{2m}\left(\sqrt{\rho(\mbf{x})}\mbf{v}(\mbf{x})\right)^2,
  \label{e11} \\
 E_{\rm int}&&=\int d^3 x
  g\left(\rho(\mbf{x})\right)^2,
  \label{e12} \\
 E_{\rm q}&&=\int d^3 x
  \dfrac{\hbar^2}{2m}\left(\nabla\sqrt{\rho(\mbf{x}}\right)^2.
  \label{e13}
\end{eqnarray}
The kinetic energy $E_{\rm kin}$ is related with the velocity field, and
$E_{\rm int}$ is the internal energy of the fluid.
The quantum energy $E_{\rm q}$ comes from the gradient of the
condensate, so that it is large when a vortex exists, as described later.
Furthermore, in order to estimate the compressibility effects, we
decompose $\sqrt{\rho}\mbf{v}$ into
$\sqrt{\rho}\mbf{v}=(\sqrt{\rho}\mbf{v})^i+(\sqrt{\rho}\mbf{v})^c$ with
$\nabla\cdot(\sqrt{\rho}\mbf{v})^i=0$. The corresponding components are
named $E_{\rm kin}^i$ and $E_{\rm kin}^c$, satisfying the relation
$E_{\rm kin}=E_{\rm kin}^i+E_{\rm kin}^c$.
The compressible kinetic
energy $E_{\rm kin}^c$ is directly concerned with the acoustic
emission.

The numerical analysis is made for the normalized GPE.
Equation (\ref{e07}) is reduced to
\begin{equation}
 i\pdfrac{f}{t}=-\nabla^2f+|f|^2f,
  \label{e14}
\end{equation}
where $f=l^{3/2}\Psi$, $x\rightarrow x/l$, $t\rightarrow t/(2ml^2/\hbar)$,
$l=8\pi a$, $f=|f|e^{i\theta}$ and
the healing length $\xi=1/\sqrt{8\pi a n_0}$.
Small perturbations around the uniform stationary solution
yield the sound wave of velocity $c=\sqrt2$.
A straight vortex is given by the stationary solution 
where $|f|$ increases from zero at the central core over the healing
length to unity at infinity.
This two dimensional solution is expressed as $f_0=|f(r)|e^{i\phi}$
in the polar coordinates $(r,\phi)$. An initial vortex 
configuration is determined by its core location and vorticity
direction. For example, a vortex with a core at $(x_0,y_0)$
in the direction of $z$ axis 
gives $f(x,y,z)=f_0(x-x_0,y-y_0)$. The initial configuration with
several vortices is given by multiplying each of the single vortex. 

To numerically integrate the GPE, we use a {\it split step
Fourier method} \cite{ssfm}. 
There are several advantages of this method:(i) 
each time 
step is split into two segments, the first of which integrates the
non-linear term in real space, and the second integrates the Laplacian
operator in Fourier space, (ii) this method is second order accurate in
the time interval of numerical analysis and all order in the space
interval of that.

\begin{figure}[p]
\begin{center}
 %WinTpicVersion2.15
\unitlength 0.1in
\begin{picture}(21.60,22.60)(1.20,-22.10)
% BOX 2 0 3 0
% 2 200 600 2200 2600
% 
\special{pn 8}%
\special{pa 200 200}%
\special{pa 2200 200}%
\special{pa 2200 2200}%
\special{pa 200 2200}%
\special{pa 200 200}%
\special{fp}%
% VECTOR 2 0 3 0
% 2 1200 2600 1200 600
% 
\special{pn 8}%
\special{pa 1200 2200}%
\special{pa 1200 200}%
\special{fp}%
\special{sh 1}%
\special{pa 1200 200}%
\special{pa 1180 267}%
\special{pa 1200 253}%
\special{pa 1220 267}%
\special{pa 1200 200}%
\special{fp}%
% VECTOR 2 0 3 0
% 2 200 1600 2200 1600
% 
\special{pn 8}%
\special{pa 200 1200}%
\special{pa 2200 1200}%
\special{fp}%
\special{sh 1}%
\special{pa 2200 1200}%
\special{pa 2133 1180}%
\special{pa 2147 1200}%
\special{pa 2133 1220}%
\special{pa 2200 1200}%
\special{fp}%
% CIRCLE 2 0 3 0
% 4 1800 1000 1800 1050 1800 1050 1800 1050
% 
\special{pn 8}%
\special{ar 1800 600 50 50  0.0000000 6.2831853}%
% CIRCLE 2 0 0 0
% 4 600 1000 600 1050 600 1050 600 1050
% 
\special{pn 8}%
\special{sh 0.600}%
\special{ar 600 600 50 50  0.0000000 6.2831853}%
% CIRCLE 2 0 3 0
% 4 600 2200 600 2250 600 2250 600 2250
% 
\special{pn 8}%
\special{ar 600 1800 50 50  0.0000000 6.2831853}%
% CIRCLE 2 0 0 0
% 4 1800 2200 1800 2250 1800 2250 1800 2250
% 
\special{pn 8}%
\special{sh 0.600}%
\special{ar 1800 1800 50 50  0.0000000 6.2831853}%
% STR 2 0 3 0
% 3 2280 1580 2280 1680 2 0
% $x$
\put(22.8000,-12.8000){\makebox(0,0)[lb]{$x$}}%
% STR 2 0 3 0
% 3 1080 420 1080 520 2 0
% $y$
\put(10.8000,-1.2000){\makebox(0,0)[lb]{$y$}}%
% STR 2 0 3 0
% 3 1080 1630 1080 1730 2 0
% $O$
\put(10.8000,-13.3000){\makebox(0,0)[lb]{$O$}}%
% STR 2 0 3 0
% 3 1530 1110 1530 1210 2 0
% $(x_0,y_0)$
\put(15.3000,-8.1000){\makebox(0,0)[lb]{$(x_0,y_0)$}}%
% STR 2 0 3 0
% 3 1520 2280 1520 2380 2 0
% $(x_0,-y_0)$
\put(15.2000,-19.8000){\makebox(0,0)[lb]{$(x_0,-y_0)$}}%
% STR 2 0 3 0
% 3 260 2280 260 2380 2 0
% $(-x_0,-y_0)$
\put(2.6000,-19.8000){\makebox(0,0)[lb]{$(-x_0,-y_0)$}}%
% STR 2 0 3 0
% 3 320 1120 320 1220 2 0
% $(-x_0,y_0)$
\put(3.2000,-8.2000){\makebox(0,0)[lb]{$(-x_0,y_0)$}}%
% STR 2 0 3 0
% 3 120 2670 120 2770 2 0
% $-L$
\put(1.2000,-23.7000){\makebox(0,0)[lb]{$-L$}}%
% STR 2 0 3 0
% 3 2050 2680 2050 2780 2 0
% $L$
\put(20.5000,-23.8000){\makebox(0,0)[lb]{$L$}}%
% STR 2 0 3 0
% 3 2250 2510 2250 2610 2 0
% $-L$
\put(22.5000,-22.1000){\makebox(0,0)[lb]{$-L$}}%
% STR 2 0 3 0
% 3 2220 610 2220 710 2 0
% $L$
\put(22.2000,-3.1000){\makebox(0,0)[lb]{$L$}}%
\end{picture}%
\end{center}
 \caption{Configuration of a single vortex on the $[-L,L]^2$ periodic
 box. When we place a vortex at $(x_0,y_0)$, to prevent the phase
 discontinuity, we add a vortex at $(-x_0,-y_0)$ and anti-vortices at
 $(-x_0,y_0)$ and $(x_0,-y_0)$.}
 \label{f02}
\end{figure}
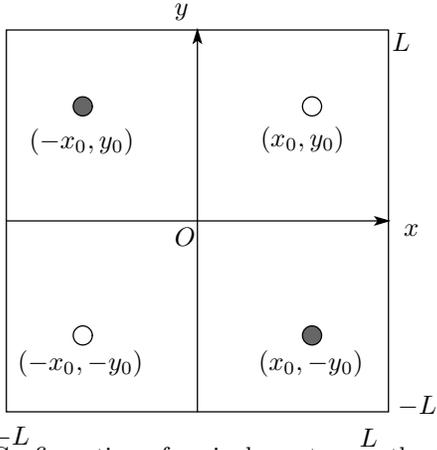

This method should take careful account of the placement of vortices.
First, we use the periodic boundary condition, because 
it allows us to use Fast-Fourier-transformation, which
saves computational times.
Secondly, for example, a single vortex configuration
in $x$-$y$ plane is shown in Fig. \ref{f02}. 
Although we need only $[0,L]\times[0,L]$ region,
the periodic box $[-L,L] \times [-L,L]$ is used in the actual numerical
calculation so that we can prevent the phase discontinuity of the
vortex. If $[0,L]\times[0,L]$ may be adopted as the periodic box, the
phase discontinuity occurs at the boundary between periodic
boxes. To prevent this discontinuity, the periodic box must be
$[-L,L]\times[-L,L]$, and we would add anti-vortices at $(-x_0,y_0)$ and
$(x_0,-y_0)$, a vortex at $(-x_0,-y_0)$.
This procedure makes the phase well-defined everywhere.

In the following figures, we usually show only
$[0,L]^3$ part for three-dimensional calculation and $[0,L]^2$ part for
two-dimensional one.
The conservation of the normalization and the total
energy is confirmed numerically. These quantities are conserved
in the order of $10^{-6}$ in all simulations.

\section{Acoustic emission in the annihilation of two anti-parallel
 vortices}\label{sound}
\begin{figure}[p]
\begin{center}
\begin{minipage}{0.47\linewidth}
\begin{center}
 \epsfxsize=\linewidth \epsfbox{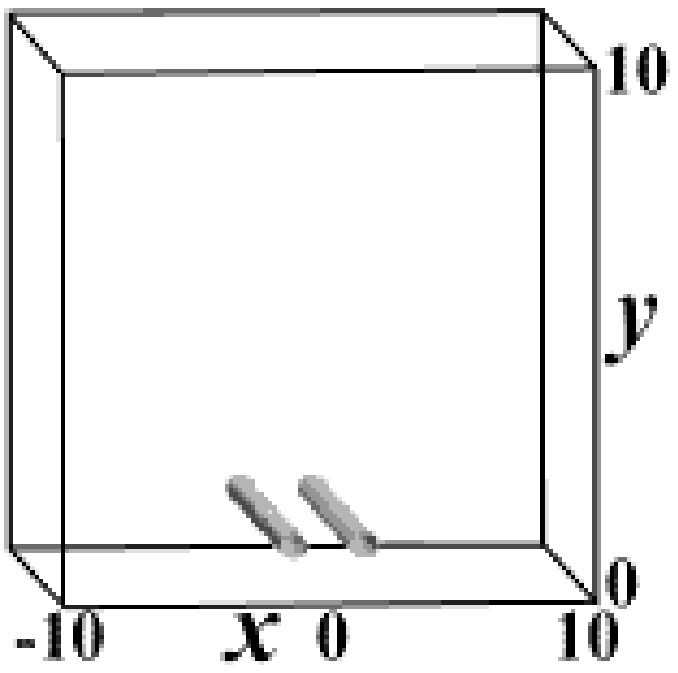}
 (a)
\end{center}
\end{minipage}
\begin{minipage}{0.47\linewidth}
\begin{center}
 \epsfxsize=\linewidth \epsfbox{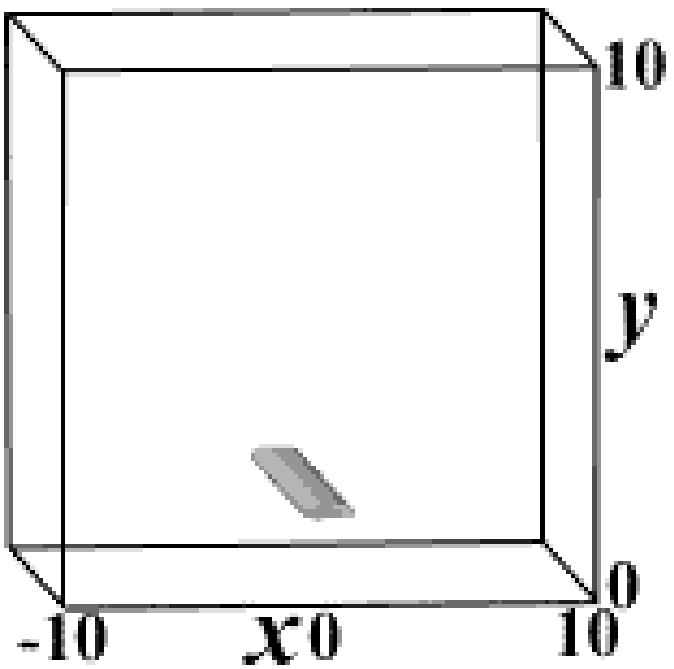}
 (b)
\end{center}
\end{minipage}
\end{center}
 \caption{Annihilation of two anti-parallel vortices, $t=0$(a) and
 $t=0.242$(b). The surface represents the contours of $|f|=0.06$.}
 \label{f06}
\end{figure}

Consider the initial configuration where two anti-parallel vortices are
placed as shown in Fig. \ref{f06}(a).
This is a simple and typical configuration, whose development includes
the Cherenkov resonance of quantized vortices studied by
Ivonin\cite{Cherenkov} and the stationary
solitary waves studied by Jones {\it et al}\cite{Jones}.
The former is deeply concerned with the acoustic emission when two
anti-parallel vortices approach and disappear.
The latter shows the analytical solutions of the GPE.
Jones {\it et. al.} investigated the axisymmetric stationary solution of
the GPE to obtain a continuous family consisting of two branches in the
momentum-energy plane. One is a vortex ring(an anti-parallel vortex
pair), and the other is a rarefaction pulse without vorticity.

As described in Appendix, two anti-parallel vortices
approach each other under the periodic boundary condition.
When the distance between them is reduced within a critical distance
$2\xi$, the velocity of two vortices becomes comparable to the sound
velocity, then the local Cherenkov resonance starts and the sound waves
are emitted\cite{Cherenkov}. The energy of the vortices is partly lost to
the sound emission, the distance between them being reduced.
These processes continue until two vortices disappear.
The sound wave emitted at the moment of the annihilation has the largest
amplitude in the process of the Cherenkov resonance.

In Fig. \ref{f06}, the initial distance between two anti-parallel
vortices, $\Delta L_0=1.5625\xi$, is less than $2\xi$,
so that they approach each other(Fig. \ref{f06}(b)).
Their singular lines of $|f|=0$ overlap for a very short
time and disappear.

\begin{figure}[p]
\begin{center}
\begin{minipage}{0.47\linewidth}
\begin{center}
 \epsfxsize=\linewidth \epsfbox{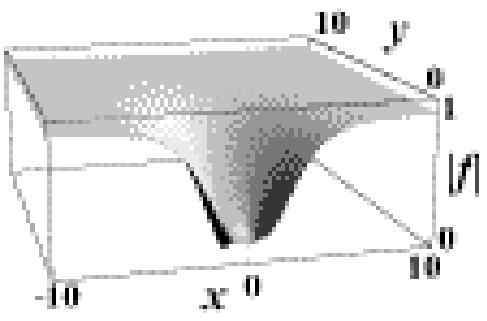}
 (a)
\end{center}
\end{minipage}
\begin{minipage}{0.47\linewidth}
\begin{center}
 \epsfxsize=\linewidth \epsfbox{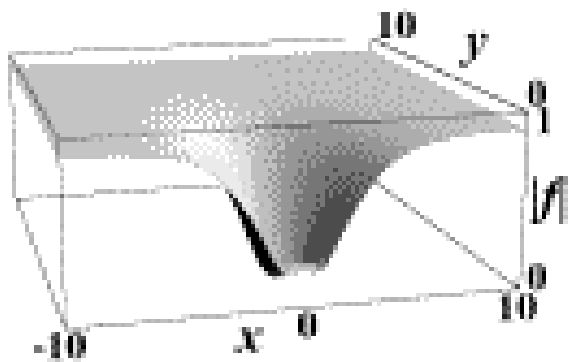}
 (b)
\end{center}
\end{minipage}
\end{center}
\begin{center}
\begin{minipage}{0.47\linewidth}
\begin{center}
 \epsfxsize=\linewidth \epsfbox{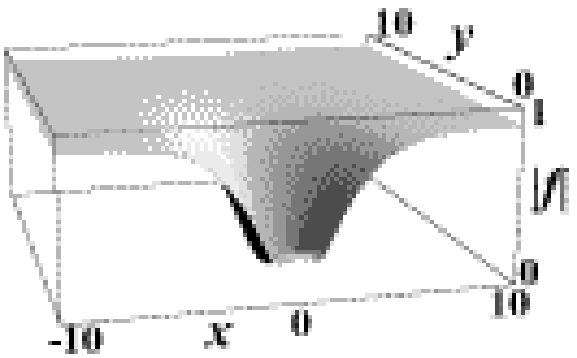}
 (c)
\end{center}
\end{minipage}
\begin{minipage}{0.47\linewidth}
\begin{center}
 \epsfxsize=\linewidth \epsfbox{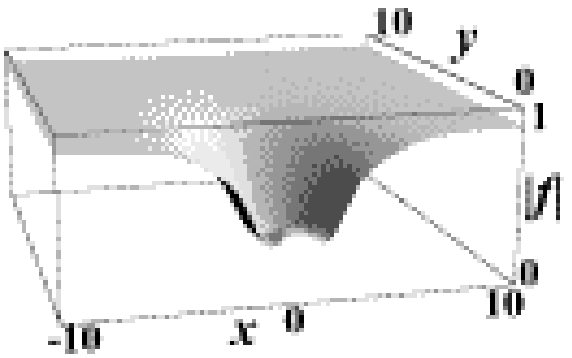}
 (d)
\end{center}
\end{minipage}
\end{center}
\begin{center}
\begin{minipage}{0.47\linewidth}
\begin{center}
 \epsfxsize=\linewidth \epsfbox{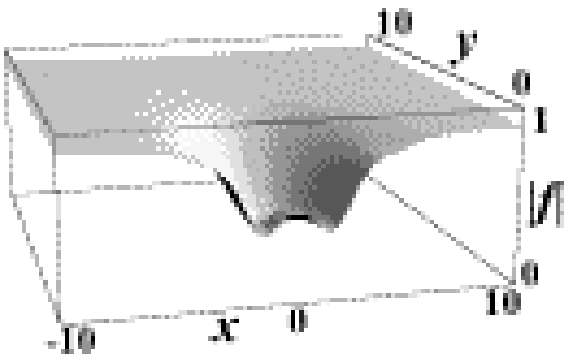}
 (e)
\end{center}
\end{minipage}
\begin{minipage}{0.47\linewidth}
\begin{center}
 \epsfxsize=\linewidth \epsfbox{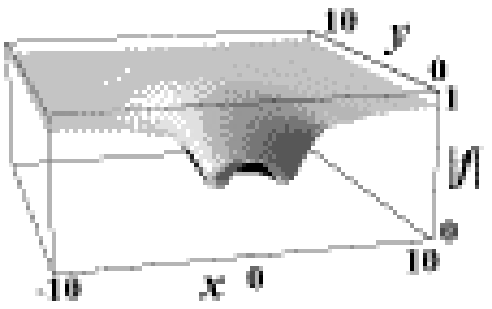}
 (f)
\end{center}
\end{minipage}
\end{center}
 \caption{Profile of $|f|$ when two anti-parallel vortices collide. The
 vertical axis refers to $|f|$, and the others represents the space
 coordinates, $t=0$(a), $t=0.242$(b), $t=0.324$(c), $t=0.61$(d),
 $t=0.814$(e), $t=1.14$(f). The depressions disappear(a)-(b), and the
 core structure are flattened gradually(c)-(f). The dimly seeable wakes
 are sound waves(f).}
 \label{f07}
\end{figure}

\begin{figure}[p]
\begin{center}
\begin{minipage}{0.47\linewidth}
\begin{center}
 \epsfxsize=\linewidth \epsfbox{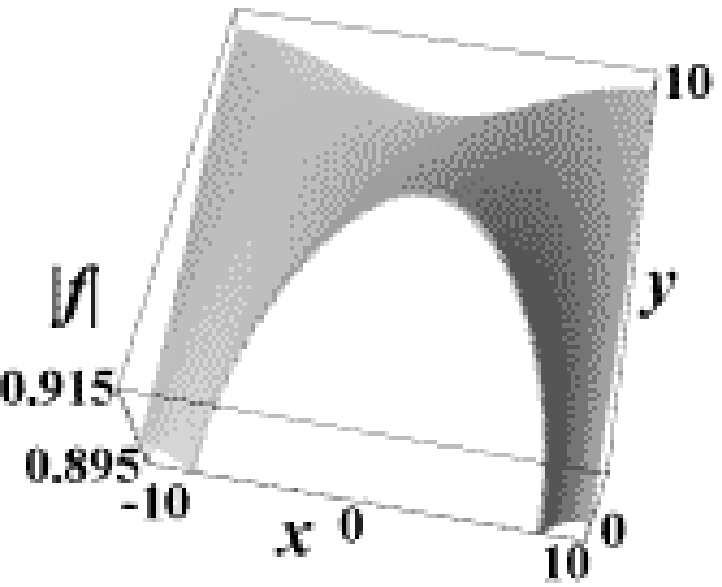}
 (a)
\end{center}
\end{minipage}
\begin{minipage}{0.47\linewidth}
\begin{center}
 \epsfxsize=\linewidth \epsfbox{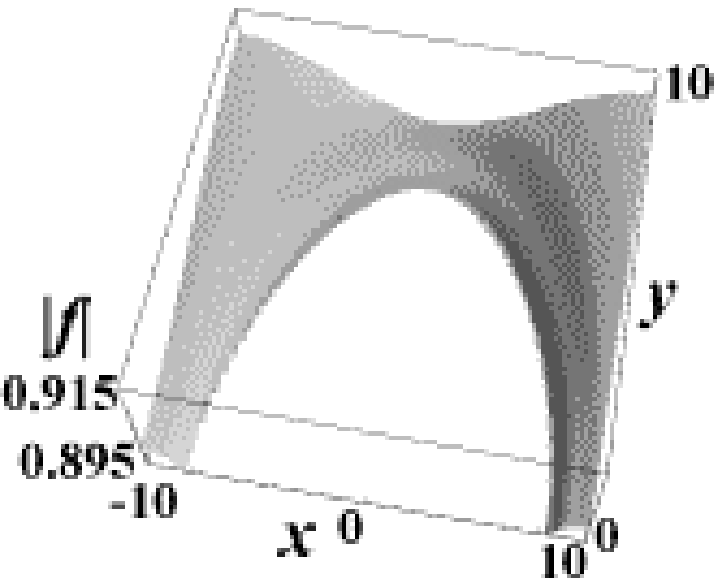}
 (b)
\end{center}
\end{minipage}
\end{center}
\begin{center}
\begin{minipage}{0.47\linewidth}
\begin{center}
 \epsfxsize=\linewidth \epsfbox{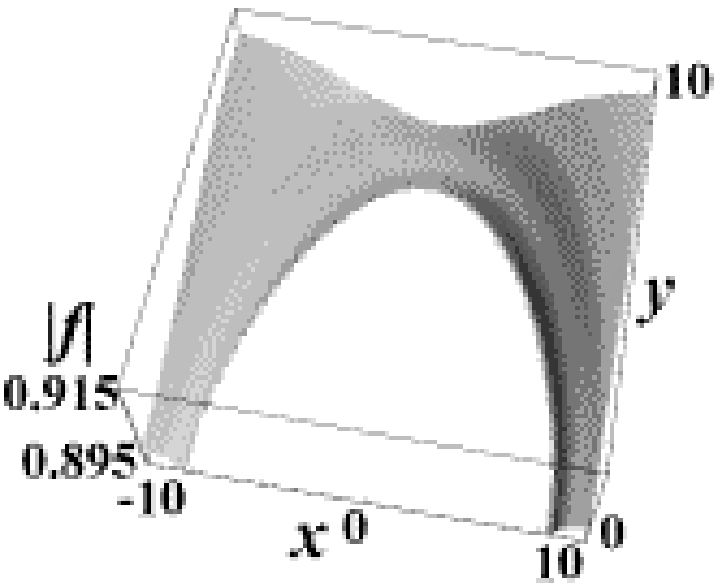}
 (c)
\end{center}
\end{minipage}
\begin{minipage}{0.47\linewidth}
\begin{center}
 \epsfxsize=\linewidth \epsfbox{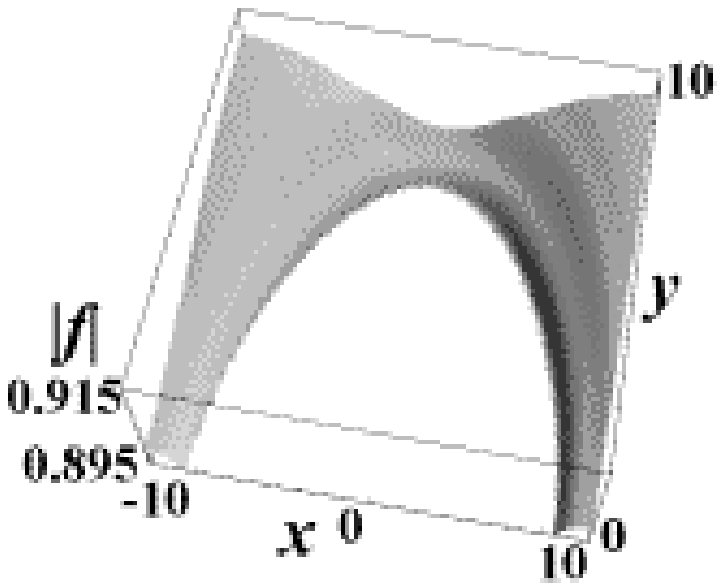}
 (d)
\end{center}
\end{minipage}
\end{center}
 \caption{Sectional profile of $|f|$. These plots show only
 $0.895\leq|f|\leq0.915$. Coordinates are the same as
 Fig. \ref{f07}, $t=0$(a), $t=1.14$(b), $t=1.18$(c), $t=1.222$(d). Some
 propagated wakes are sound waves.}
 \label{f08}
\end{figure}

We consider the change of the condensate density $|f|$ in the process
of Fig. \ref{f06}. Note that
it is necessary to separate 
carefully the effect of acoustics from that due to vortex motion. In
order to reveal both the vortex motion and the acoustic emission, we
show the change of $|f|$ in Fig. \ref{f07}.
Initially two very close depressions represent the singular cores of
the vortices((a)).
They become close and overlap for a
very short time and disappear((b)). After their annihilation
at $t\simeq0.2$, the vestigial of two vortex cores becomes flattened
gradually((c)-(f)).
%%%%%%%%%%%%%%%%%%%%%%%%%%%%%%%%%
The sound propagations are dimly shown in Fig. \ref{f07};
in order to show them more clearly, Fig. \ref{f08} shows
the sectional profile of the density $|f|$.
%%%%%%%%%%%%%%%%%%%%%%%%%%%%%%%%%
After the annihilation,
some wakes are found to propagate outside over the global profile; this is
nothing but a sound wave. Its propagation velocity is of the same order
as $c=\sqrt2$.

The motion of the vestigial of two vortex cores is related with the stationary
solitary wave studied by Jones {\it et al}\cite{Jones}.
We investigate the contour plots of the condensate density.
These plots resemble those obtained by Jones {\it et al}, although they
studied the stationary solution.
Our dynamical solution follows approximately the vortex solution, makes
two vortex cores disappear, then follows the rarefaction pulse
solution. The vestigial of the vortices correspond to the rarefaction
pulse.
The above sound waves propagate over the profile of the rarefaction pulse.

\begin{figure}[p]
\begin{center}
 \epsfxsize=\linewidth \epsfbox{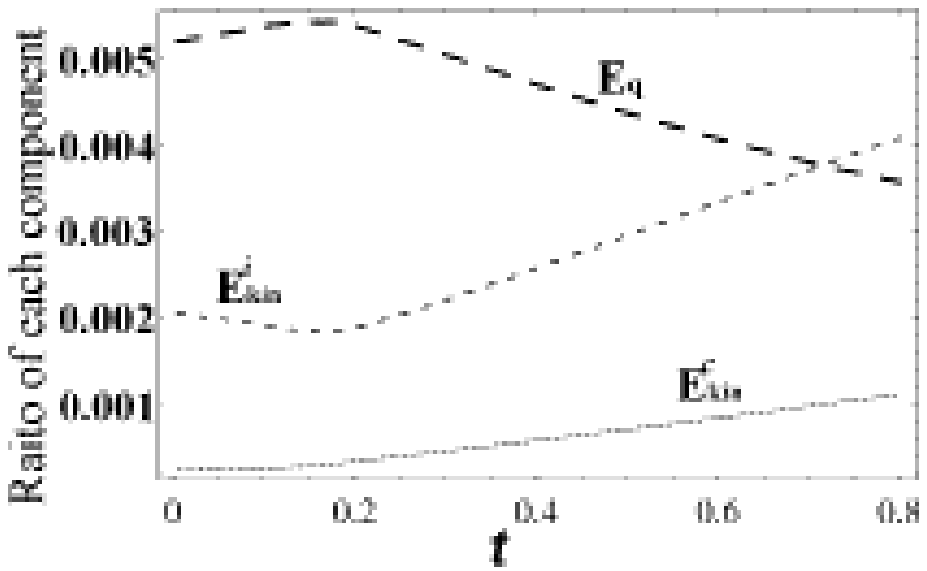}
\end{center}
 \caption{Change in each energy component for the process of
 Fig. \ref{f06}. The thick broken line , the thin broken line and the
 dotted line shoe $E_{\rm q}/E_{\rm tot}$, $E_{\rm kin}^i/E_{\rm tot}$
 and $E_{\rm kin}^c/E_{\rm tot}$.}
 \label{f09}
\end{figure}

The change in each energy component is shown in Fig .\ref{f09}.
This figure does not show the internal energy component $E_{\rm int}$,
because it is much larger than the other components,
$E_{\rm int}/E_{\rm total}\sim 1$.
The compressible and incompressible energy increases after two
vortices vanish at $t\simeq0.2$. The increase in the
compressible energy is reasonable.
On the other hand, the increase in the incompressible
energy may remain controversial.
The quantum energy of the vortex cores is transferred
irreversibly into the kinetic energy as described later.

\begin{figure}[p]
\begin{center}
\begin{minipage}{0.47\linewidth}
\begin{center}
 \epsfxsize=\linewidth \epsfbox{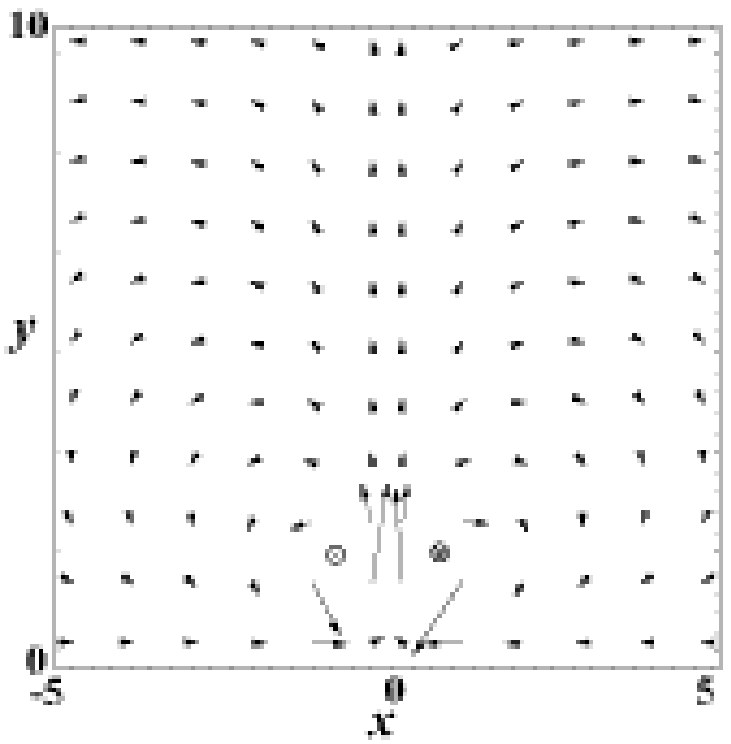}
 (a)
\end{center}
\end{minipage}
\begin{minipage}{0.47\linewidth}
\begin{center}
 \epsfxsize=\linewidth \epsfbox{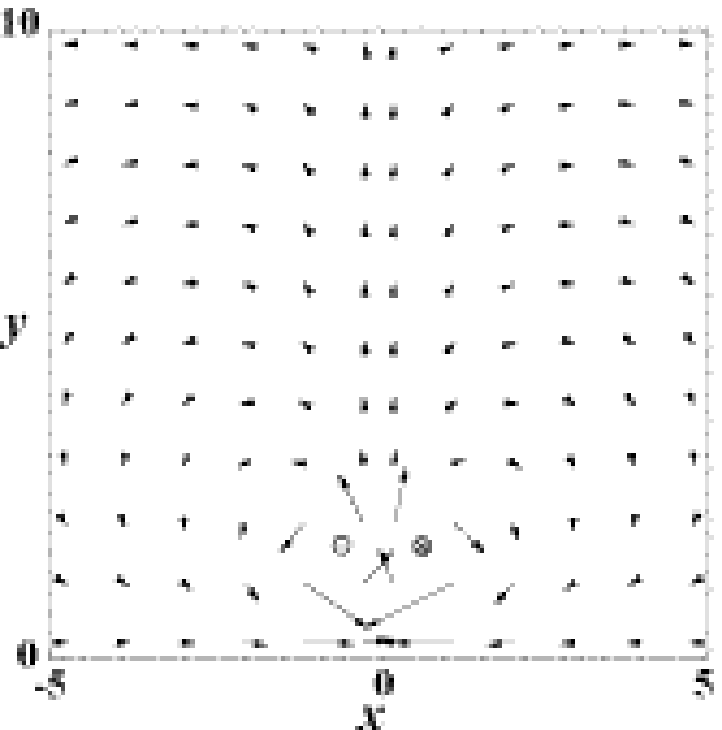}
 (b)
\end{center}
\end{minipage}
\end{center}
\begin{center}
\begin{minipage}{0.47\linewidth}
\begin{center}
 \epsfxsize=\linewidth \epsfbox{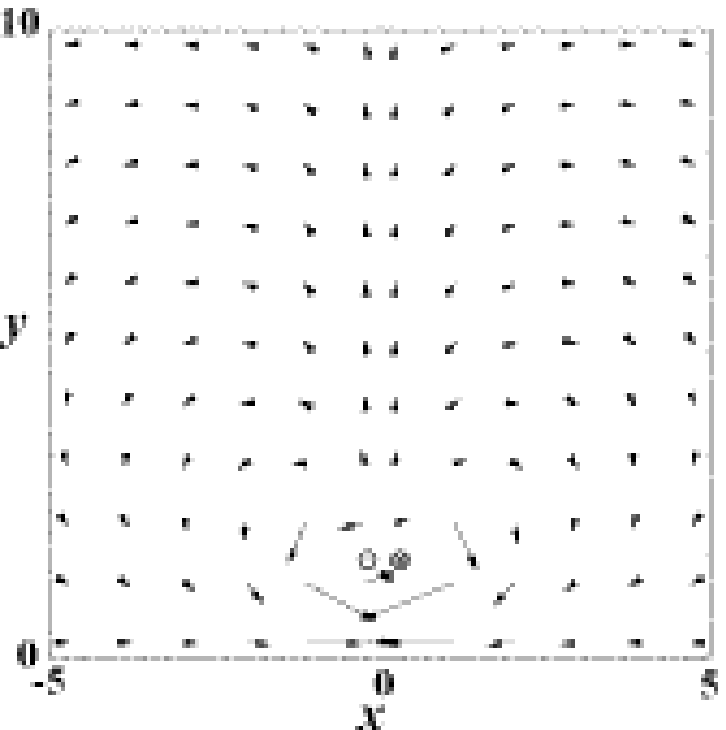}
 (c)
\end{center}
\end{minipage}
\begin{minipage}{0.47\linewidth}
\begin{center}
 \epsfxsize=\linewidth \epsfbox{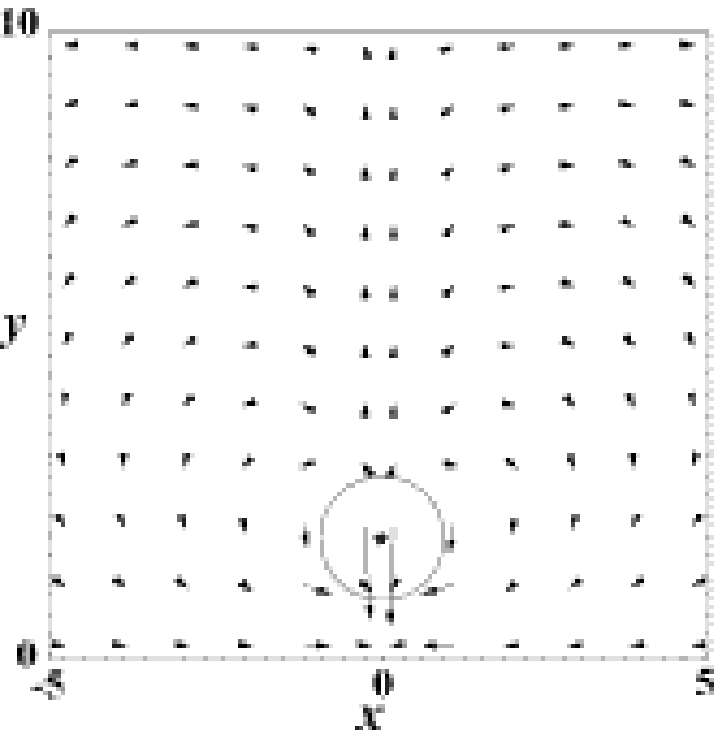}
 (d)
\end{center}
\end{minipage}
\end{center}
\begin{center}
\begin{minipage}{0.47\linewidth}
\begin{center}
 \epsfxsize=\linewidth \epsfbox{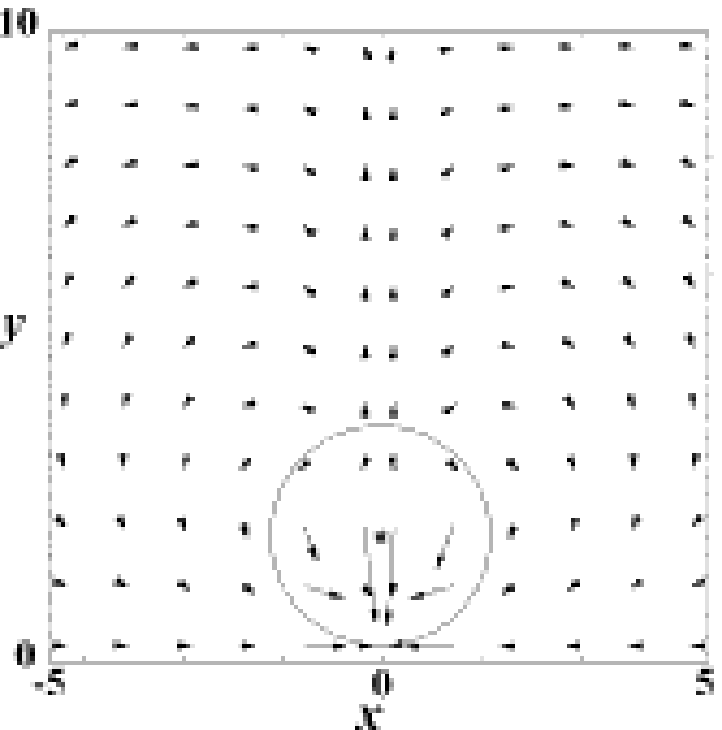}
 (e)
\end{center}
\end{minipage}
\begin{minipage}{0.47\linewidth}
\begin{center}
 \epsfxsize=\linewidth \epsfbox{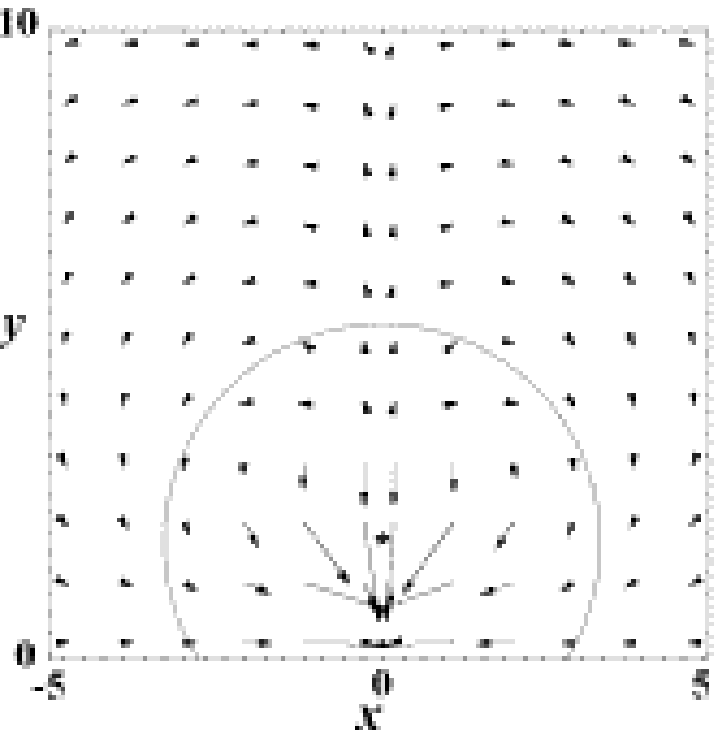}
 (f)
\end{center}
\end{minipage}
\end{center}
\begin{center}
\begin{minipage}{0.47\linewidth}
\begin{center}
 \epsfxsize=\linewidth \epsfbox{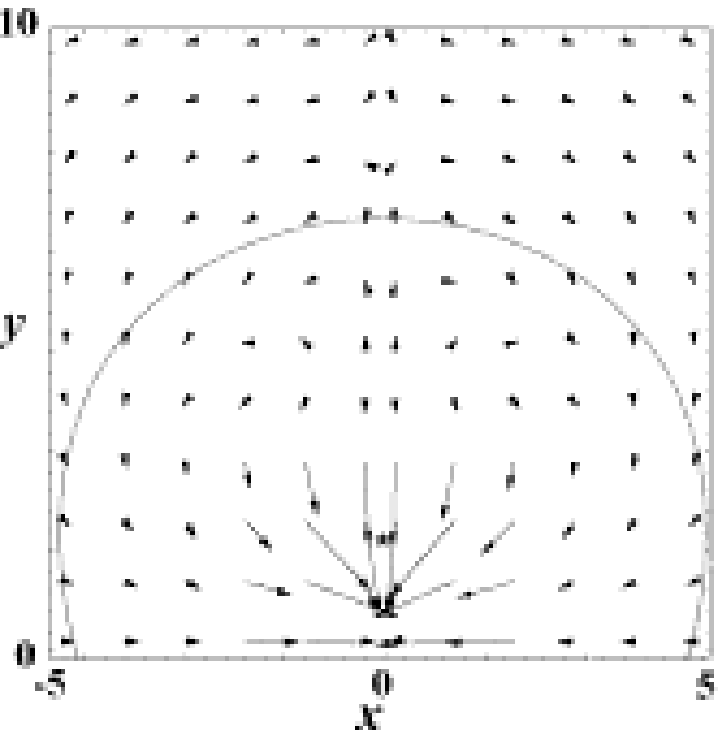}
 (g)
\end{center}
\end{minipage}
\begin{minipage}{0.47\linewidth}
\begin{center}
 \epsfxsize=\linewidth \epsfbox{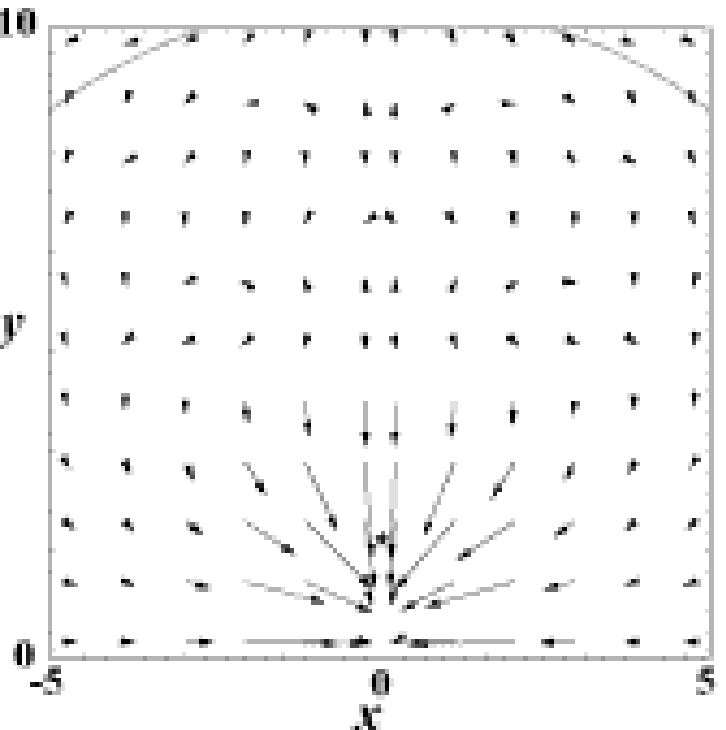}
 (h)
\end{center}
\end{minipage}
\end{center}
 \caption{Vector field $\nabla\theta$, t=0(a), t=0.14(b),
 t=0.192(c), t=0.316(d), t=0.406(e), t=0.61(f), t=0.814(g),
 t=1.222(h). The symbols $\odot$ and $\otimes$ denote a vortex and an
 anti-vortex,
 respectively. The vector heads denote the direction of $\nabla\theta$,
 and its length denotes $|\nabla\theta|$. The symbol $\star$ shows the
 annihilation point, and the arc shows the crest of the sound wave
 emitted at the annihilation. For convenience of visualization, the
 vector length is stretched.}
 \label{f10}
\end{figure}

The sound wave affects not only the density $|f|$ but also the field of
the phase $\theta$. The vector field $\mbf{v}_s=\nabla \theta$ in
$x$-$y$ planes are shown in
Fig. \ref{f10}. The symbols $\odot$ and $\otimes$ denote the singular
lines with $|f|=0$ of a vortex and an anti-vortex, respectively.
Two vortices become close((a)-(c)) and disappear((d)). Before their
annihilation , they become close together with the circulative
field around itself.
Just before the annihilation, $|\mbf{v}_s|$ between two vortices
grows very much but the density $|f|$ there is reduced.
When their singular lines overlap, $\mbf{v}_s$ between them disappears,
while the circulative field around them still remains((d)). 
Since the propagation speed of the event, {\it i.e.}
``the annihilation of the singular lines'', is of the same order as $c$,
%and the distant $\mbf{v}_s$ from the annihilation point is very small,
$\mbf{v}_s$ far apart from the annihilation point keeps the circulative
flow for a certain time((d)-(f)). After this event information comes up,
the circulative flow disappears((g)-(h)).
The arc shown in Fig. \ref{f10} indicates the crest of the sound wave
emitted at the annihilation, which is estimated by the density
propagation as shown in Fig. \ref{f07} and \ref{f08}.
The arc shows more clearly the disappearing
processes of the circulative flow. As described before,
the quantum energy of the vestigial of the singular cores
are reduced and transferred into the kinetic energy. The increase in the
kinetic energy results from the large velocity field near the
annihilation point shown in Fig. \ref{f10}(f)-(h).

\begin{figure}[p]
\begin{center}
\begin{minipage}{0.47\linewidth}
\begin{center}
 \epsfxsize=\linewidth \epsfbox{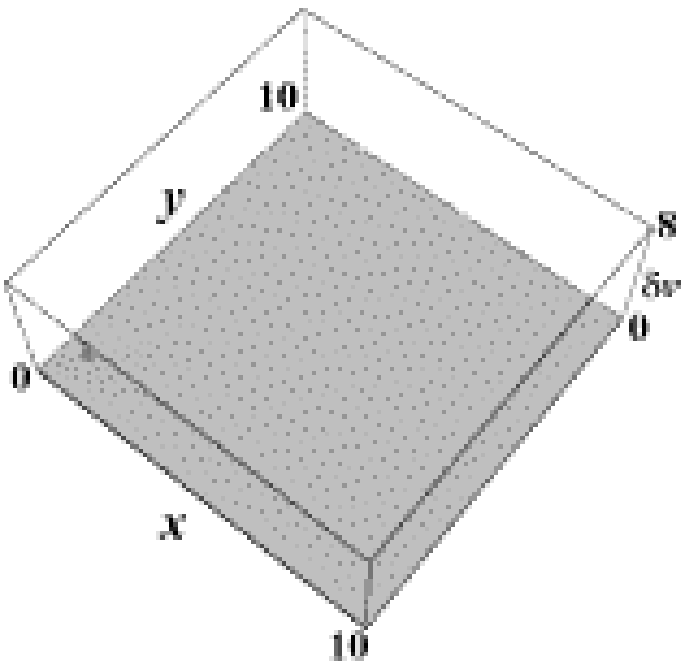}
 (a)
\end{center}
\end{minipage}
\begin{minipage}{0.47\linewidth}
\begin{center}
 \epsfxsize=\linewidth \epsfbox{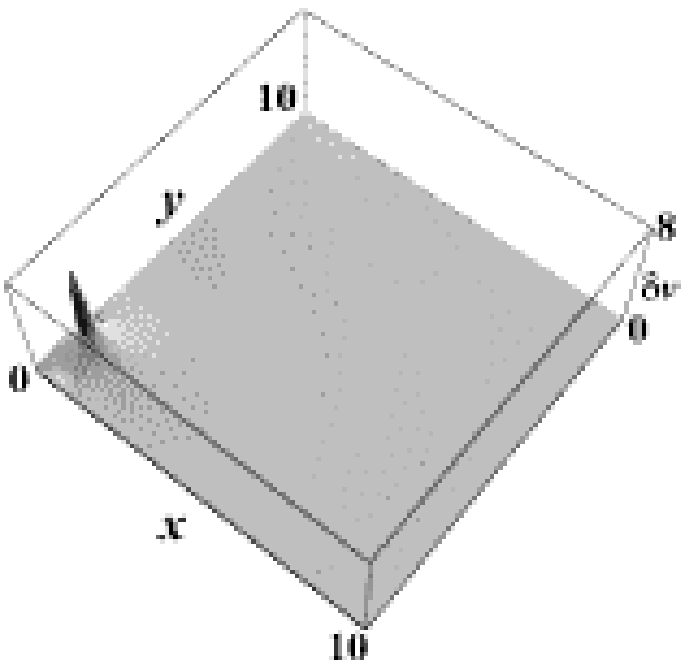}
 (b)
\end{center}
\end{minipage}
\end{center}
\begin{center}
\begin{minipage}{0.47\linewidth}
\begin{center}
 \epsfxsize=\linewidth \epsfbox{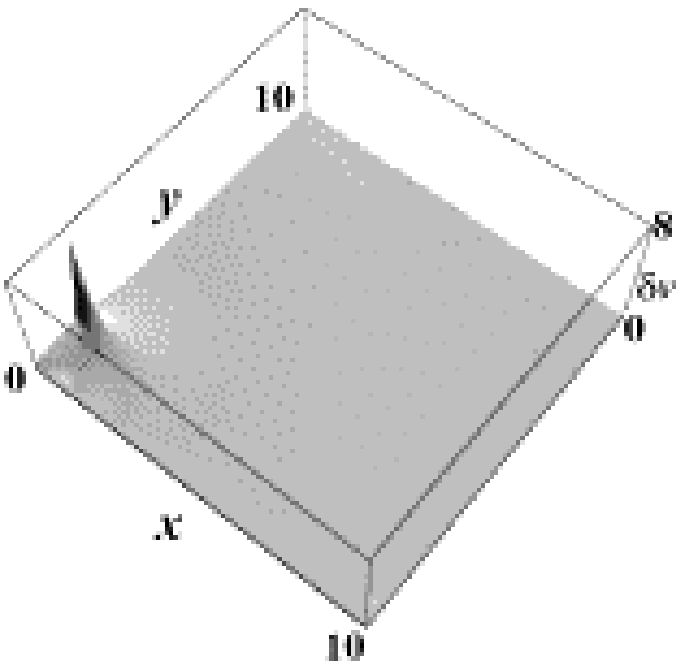}
 (c)
\end{center}
\end{minipage}
\begin{minipage}{0.47\linewidth}
\begin{center}
 \epsfxsize=\linewidth \epsfbox{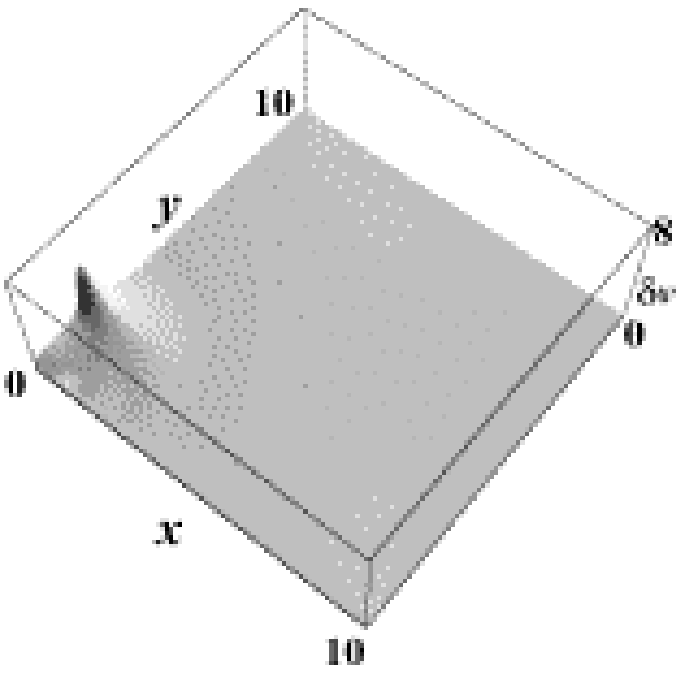}
 (d)
\end{center}
\end{minipage}
\end{center}
\begin{center}
\begin{minipage}{0.47\linewidth}
\begin{center}
 \epsfxsize=\linewidth \epsfbox{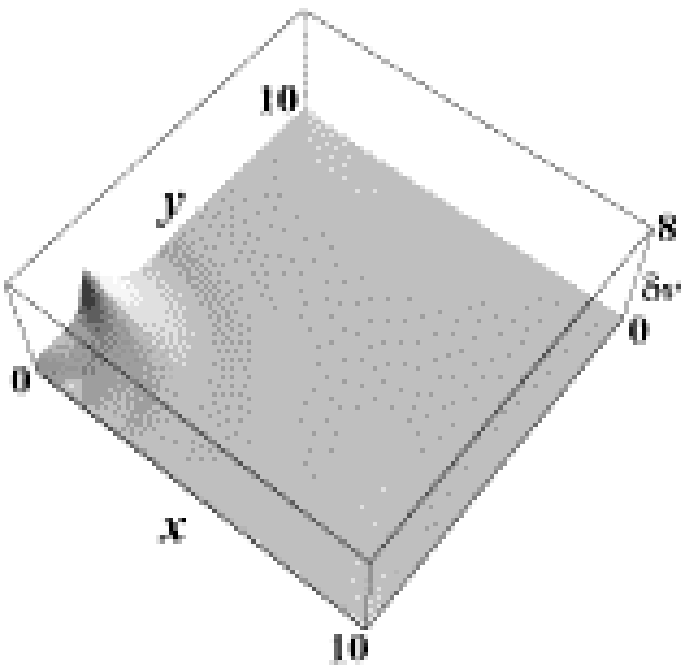}
 (e)
\end{center}
\end{minipage}
\begin{minipage}{0.47\linewidth}
\begin{center}
 \epsfxsize=\linewidth \epsfbox{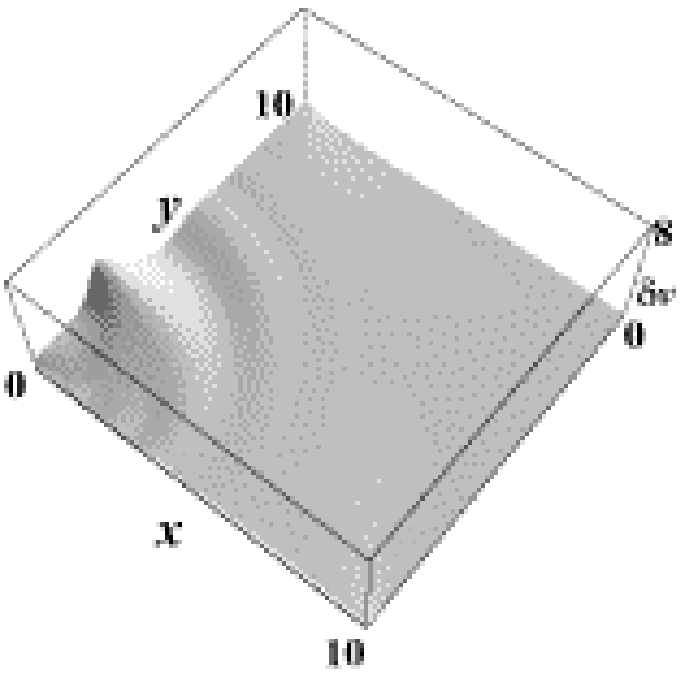}
 (f)
\end{center}
\end{minipage}
\end{center}
\begin{center}
\begin{minipage}{0.47\linewidth}
\begin{center}
 \epsfxsize=\linewidth \epsfbox{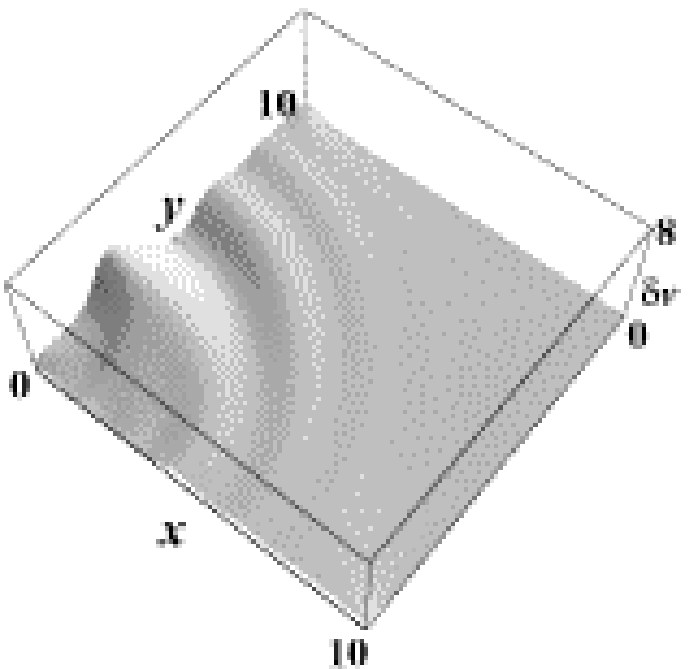}
 (g)
\end{center}
\end{minipage}
\begin{minipage}{0.47\linewidth}
\begin{center}
 \epsfxsize=\linewidth \epsfbox{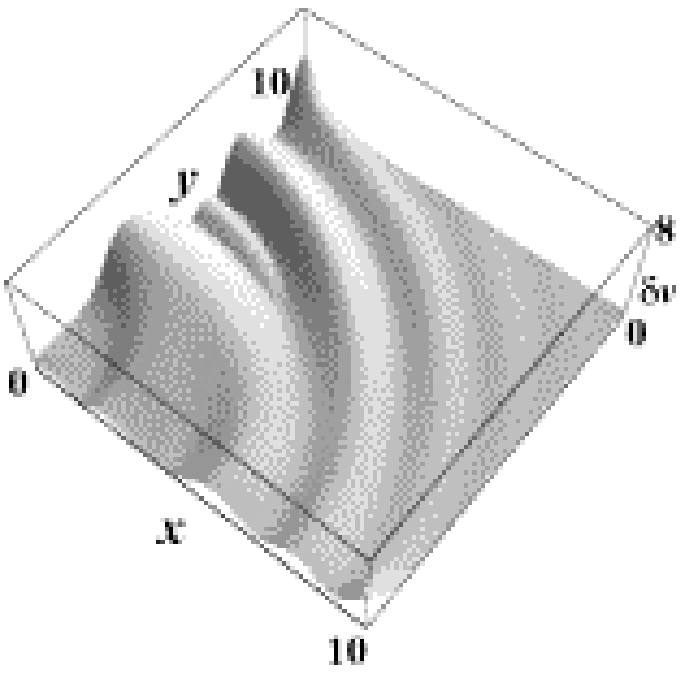}
 (h)
\end{center}
\end{minipage}
\end{center}
 \caption{Change $\delta v(x,y,t)$ of the velocity field
 from the initial state, t=0(a), t=0.14(b), t=0.192(c), t=0.316(d),
 t=0.406(e), t=0.61(f), t=0.814(g), t=1.222(h).}
 \label{f061}
\end{figure}

Figure \ref{f061} shows the change of the vector field
from the initial state
\begin{equation}
 \delta v(x,y,t)=\dfrac{\left|\mbf{v}(x,y,t)-\mbf{v}(x,y,0)\right|}
 {\left|\mbf{v}(x,y,0)\right|}.
 \label{e061}
\end{equation}
These figures show the propagations of the change of the vector field,
reflecting to the propagation of the sound waves as described above.
The distance between two anti-parallel vortices oscillates by the
acoustic emission of the Cherenkov resonance\cite{Cherenkov},
including the contribution from the vortex motion too.
Although the circulative flow is slightly disarranged by
the sound wave by the Cherenkov resonance for a moment,
then it recovers the previous state((f)-(g)) mostly.
This recovering mechanism breaks down gradually as the amplitude of the
sound wave increases, eventually the circulative flow disappears after
the maximum sound wave emitted at the annihilation passes by.

%%%%%%%%%%%%%%%%%%%%%%%%%%%%%%%%%%%%%%%%%%%%%%%%%%%%
%%%%%%%%%%%%%%%%%%%%%%%%%%%%%%%%%%%%%%%%%%%%%%%%%%%%
%%%%%%%%%%%%%%%%%%%%%%%%%%%%%%%%%%%%%%%%%%%%%%%%%%%%

We investigate also the distribution of the incompressible
kinetic energy density
\begin{equation}
 e_{\rm kin}^i(\mbf{x})=\left(\left(\sqrt{\rho}\mbf{v}\right)^i\right)^2
 \label{den_eki}
\end{equation}
and the compressible kinetic energy density
\begin{equation}
 e_{\rm kin}^c(\mbf{x})=\left(\left(\sqrt{\rho}\mbf{v}\right)^c\right)^2.
 \label{den_ekc}
\end{equation}
The distribution of $e_{\rm kin}^c(\mbf{x})$
has small peaks
on the sound waves and is also spread into the whole system.
On the other hand, that of $e_{\rm kin}^c(\mbf{x})$ has a very large
peak near the vestigial of the singular cores,
which contributes to 
the increasing in the incompressible energy in Fig. \ref{f09}.
Eventually, because there is no dissipative mechanism,
these energy components are mixed and oscillates after that.

\section{Detailed process of reconnection}\label{rec1}
\begin{figure}[p]
\begin{center}
\begin{minipage}{0.47\linewidth}
\begin{center}
 \epsfxsize=\linewidth \epsfbox{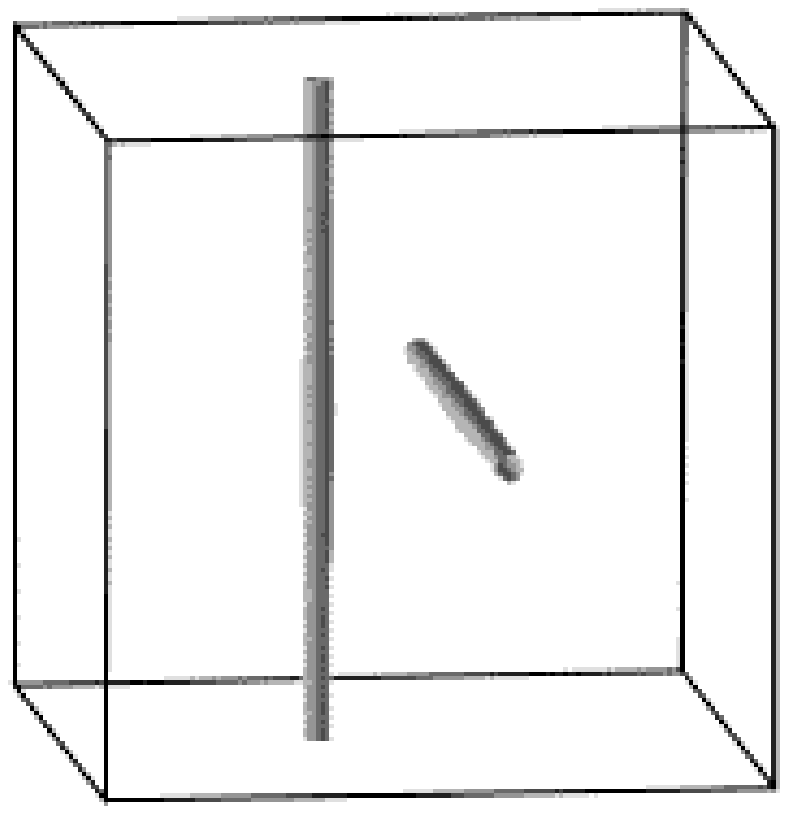}
 (a)
\end{center}
\end{minipage}
\begin{minipage}{0.47\linewidth}
\begin{center}
 \epsfxsize=\linewidth \epsfbox{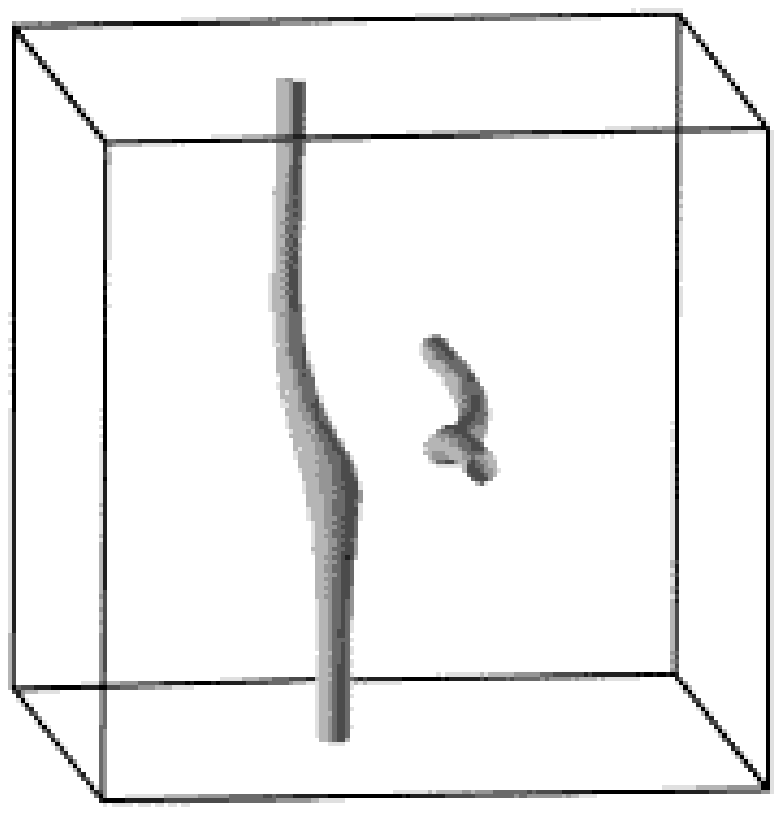}
 (b)
\end{center}
\end{minipage}
\end{center}
\begin{center}
\begin{minipage}{0.47\linewidth}
\begin{center}
 \epsfxsize=\linewidth \epsfbox{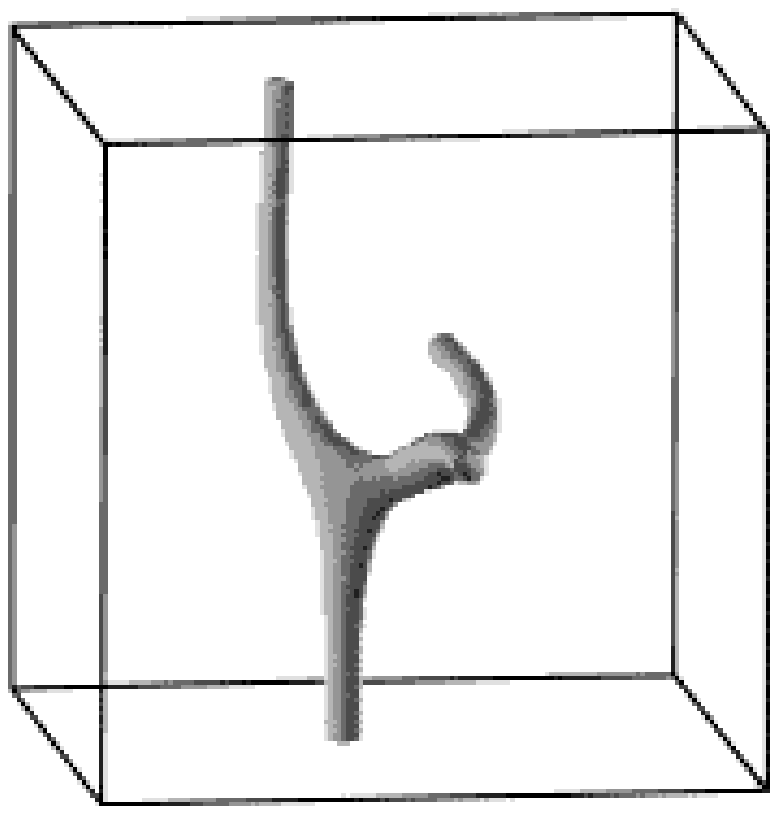}
 (c)
\end{center}
\end{minipage}
\begin{minipage}{0.47\linewidth}
\begin{center}
 \epsfxsize=\linewidth \epsfbox{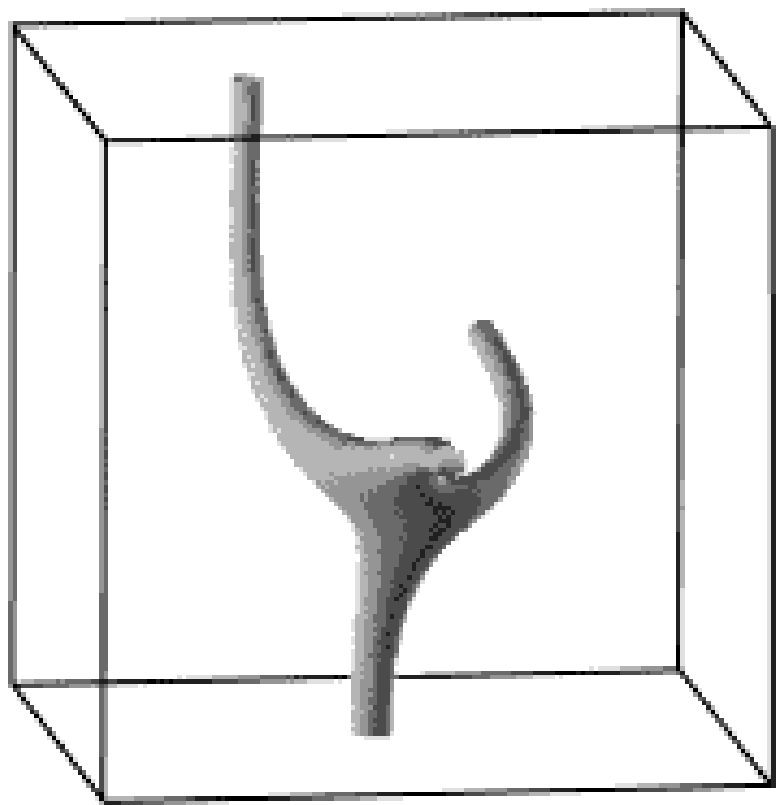}
 (d)
\end{center}
\end{minipage}
\end{center}
\begin{center}
\begin{minipage}{0.47\linewidth}
\begin{center}
 \epsfxsize=\linewidth \epsfbox{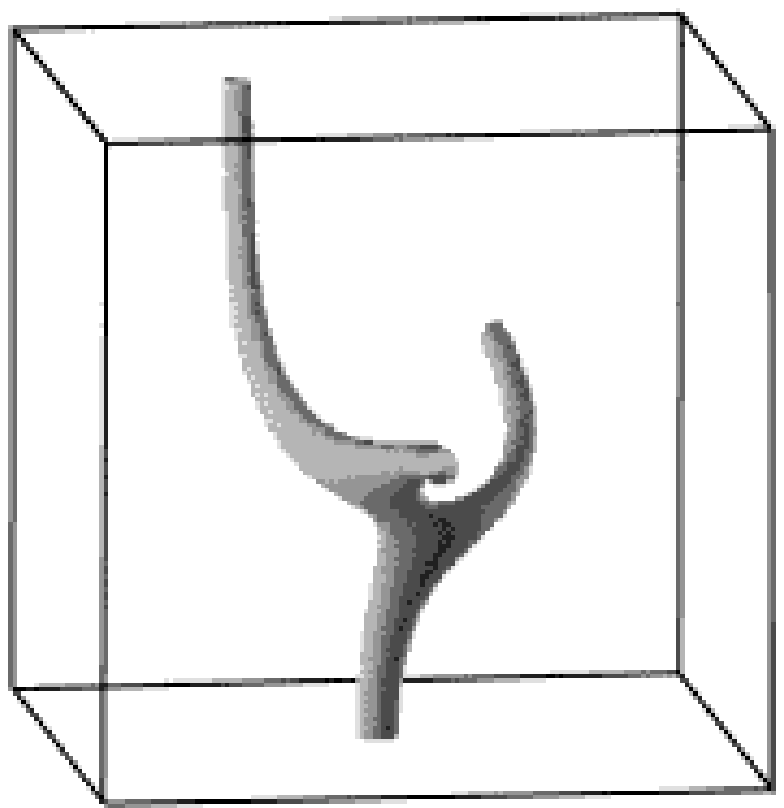}
 (e)
\end{center}
\end{minipage}
\begin{minipage}{0.47\linewidth}
\begin{center}
 \epsfxsize=\linewidth \epsfbox{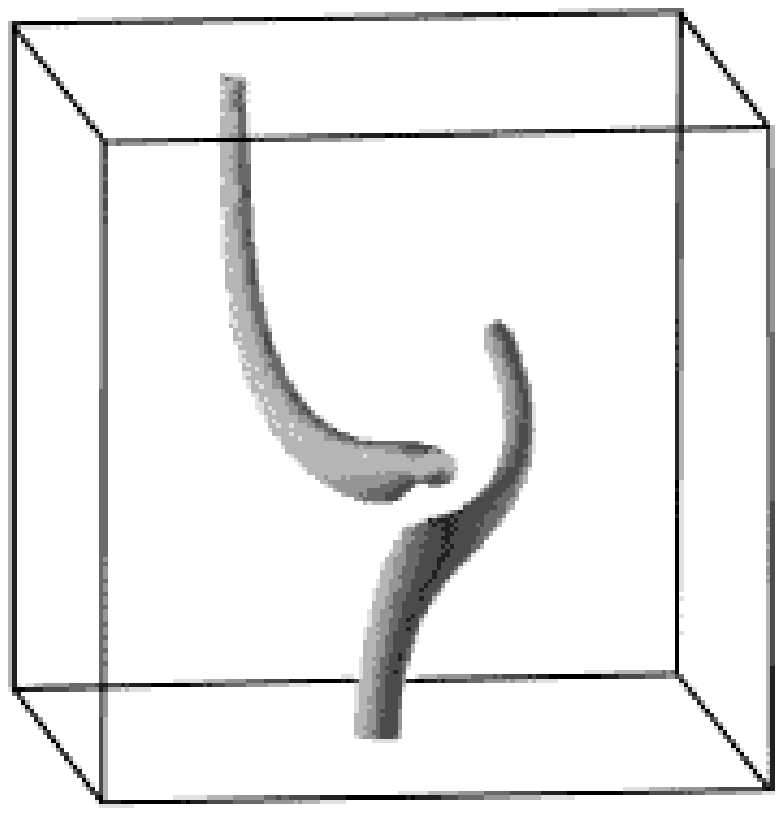}
 (f)
\end{center}
\end{minipage}
\end{center}
 \caption{Reconnection of two vortices, $t=0$(a), $0.816$(b),
 $1.224$(c), $2.448$(d), $2.856$(e), $3.06$(f). The surface represents
 the contour of $|f|=0.1$. Two vortices cause local twists in each
 other(b) so that they become anti-parallel(c), then reconnect(d) and
 leave separately(e),(f).}
 \label{f01}
\end{figure}

\begin{figure}[p]
\begin{center}
\begin{minipage}{0.47\linewidth}
\begin{center}
 \epsfxsize=\linewidth \epsfbox{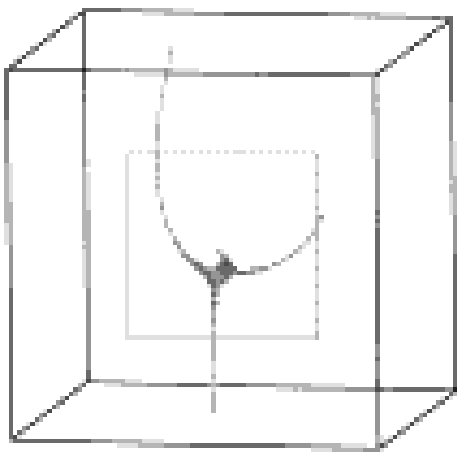}
 (a)
\end{center}
\end{minipage}
\begin{minipage}{0.47\linewidth}
\begin{center}
 \epsfxsize=\linewidth \epsfbox{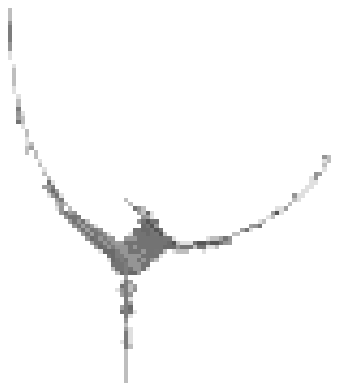}
 (b)
\end{center}
\end{minipage}
\end{center}
\begin{center}
\begin{minipage}{0.47\linewidth}
\begin{center}
 \epsfxsize=\linewidth \epsfbox{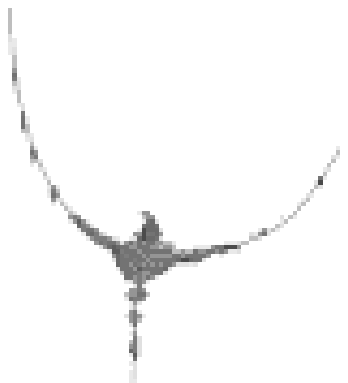}
 (c)
\end{center}
\end{minipage}
\begin{minipage}{0.47\linewidth}
\begin{center}
 \epsfxsize=\linewidth \epsfbox{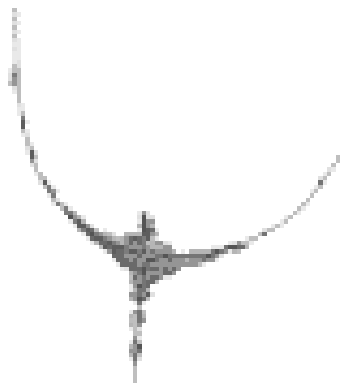}
 (d)
\end{center}
\end{minipage}
\end{center}
 \caption{More detailed reconnection process of two vortices,
 $t=1.386$(a,b), $1.428$(c), $1.46$(d). The surface represents the
 contour of $|f|=0.03$. Two lines show the singular line by tracing
 the local minimum density less than $|f|\sim0.06$.}
 \label{ff1}
\end{figure}

In this section, to show the detailed processes of the reconnection,
we study the three-dimensional dynamics starting from the configuration
where two vortices are placed at
a right angle as shown in Fig .\ref{f01}(a). The grayed surfaces
represents the contour of $|f|=0.1$, which is more inside in the core
than that of $|f|=\sqrt{0.3}$ 
calculated by Koplik {\it et al}\cite{Koplik}. Two vortices cause local
twists in each other((b)) so that they become locally anti-parallel at
the closest point((c)) and approach each other, then reconnect((d)) and
leave separately((e)-(f)). This motion is the same as that
found by the filament calculation\cite{Schwarz}.
The locally anti-parallel vortices which are closer than the critical
distance become to follow the scenario of the Cherenkov resonance
described in Sec. \ref{sound}: the anti-parallel part moves locally with
the velocity comparable to the sound velocity, emits the sound wave and
crosses at a point, corresponding to the two-dimensional annihilation.
In order to investigate the switching of the singular core, we check the
motion of the surface of $|f|=0.03$ too as shown in Fig. \ref{ff1}.
%%%%%%%%%%%%%%%%%%%%%%%%%%%%%%%%%%%%%%%%
Two lines in Fig. \ref{ff1}(a)-(d) show
%%%%%%%%%%%%%%%%%%%%%%%%%%%%%%%%%%%%%%%%
the singular line of $|f|\simeq 0$.
These lines reconnect at a point, {\it i.e.} the reconnecting point.
Two vortices are found to reconnect at $t\sim1.45$.
The detailed processes of the reconnection are as follows:
the singular cores of two vortices cross once, and keep the transient
crossing
for a very short time, path through each other, eventually leaving by
changing their topology.
The quantized vortices reconnect without going through the classical 
bridge-like reconnection processes. The quantized vortex
reconnection is not contrary to the Kelvin's circulation theorem,
as discussed in Sec. \ref{kel}.
%%%%%%%%%%%%%%%%%%%%%%%%%%%%%%%%%%%%%%%%%%%%%%%%%%%
%Koplik {\it et al.} reported that the surface of $|f|=\sqrt{0.3}$ makes
%a transitional bridge-like structure \cite{Koplik}, while we find
%that the thinner core does not go through such a state. 
%%%%%%%%%%%%%%%%%%%%%%%%%%%%%%%%%%%%%%%%%%%%%%%%%%%

\begin{figure}[p]
\begin{center}
 \epsfxsize=\linewidth \epsfbox{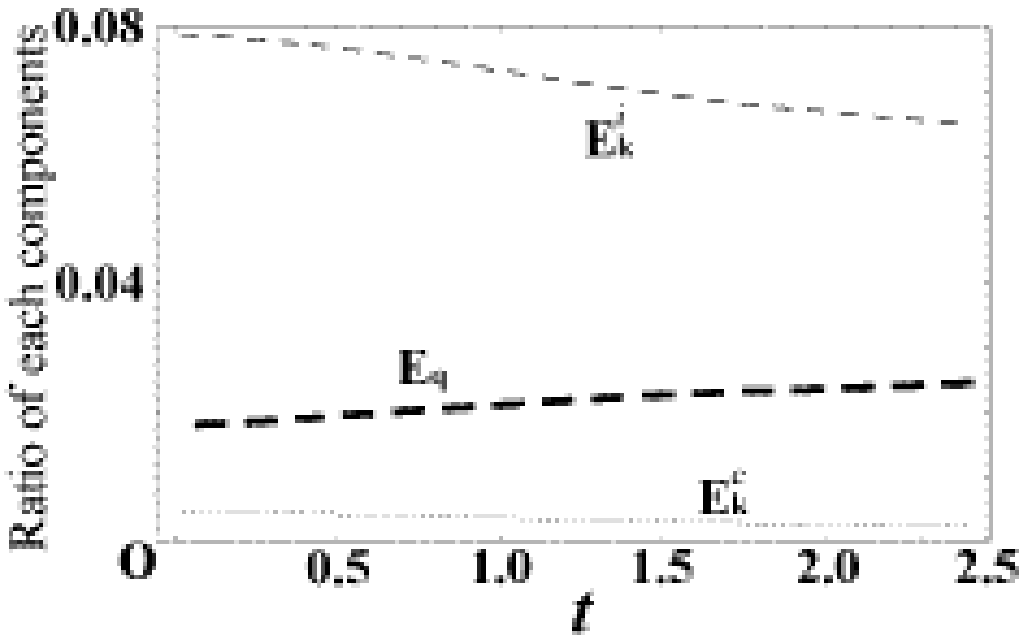}
\end{center}
 \caption{Change in the energy component for the process of
 Fig. \ref{f01}. The thick broken line ,the thin broken line and the
 dotted line show respectively
 $E_{\rm q}/E_{\rm tot},E_{\rm kin}^i/E_{\rm tot}$
 and $E_{\rm kin}^i/E_{\rm tot}$.}
 \label{f04}
\end{figure}

Figure \ref{f04} shows the changes of the ratio of each energy component
of the process of Fig. \ref{f01},which does not show $E_{\rm int}$
by the same reason in Sec. \ref{sound}.
During the processes of the reconnection,
$E_{\rm kin}^c$ and $E_{\rm kin}^i$ are reduced,
and $E_{\rm q}$ increases.
If there would be the concurrent acoustic emission,
the compressible component should increase.
The change of the energy components will be discussed in detail in
Sec. \ref{energy}.

\section{Sequential processes of reconnection and concurrent
acoustic emission}\label{rec2}
\begin{figure}[p]
 \begin{center}
  \epsfxsize=0.9\linewidth \epsfbox{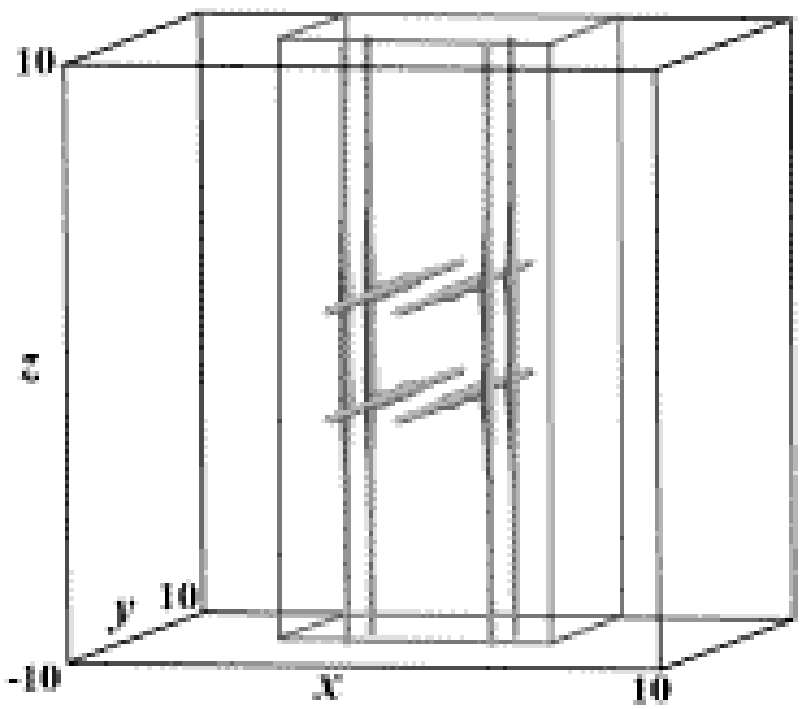}

  (a)
 \end{center}
 \begin{center}
  \begin{minipage}{0.47\linewidth}
   \begin{center}
    \epsfxsize=\linewidth \epsfbox{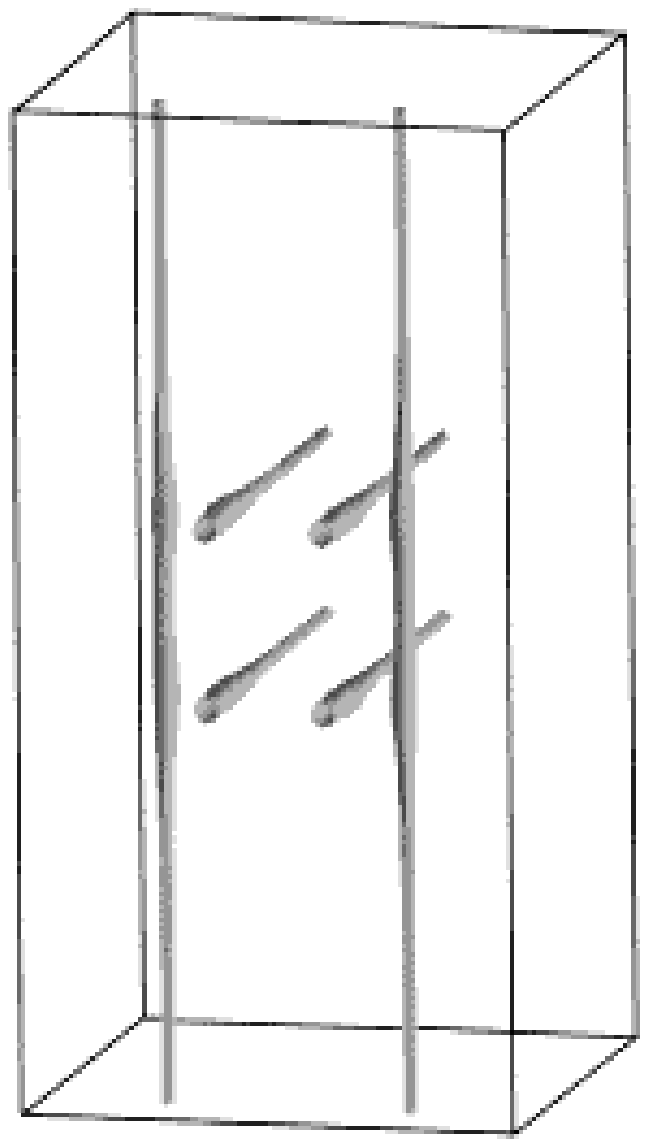}
    (b)
   \end{center}
  \end{minipage}
  \begin{minipage}{0.47\linewidth}
   \begin{center}
    \epsfxsize=\linewidth \epsfbox{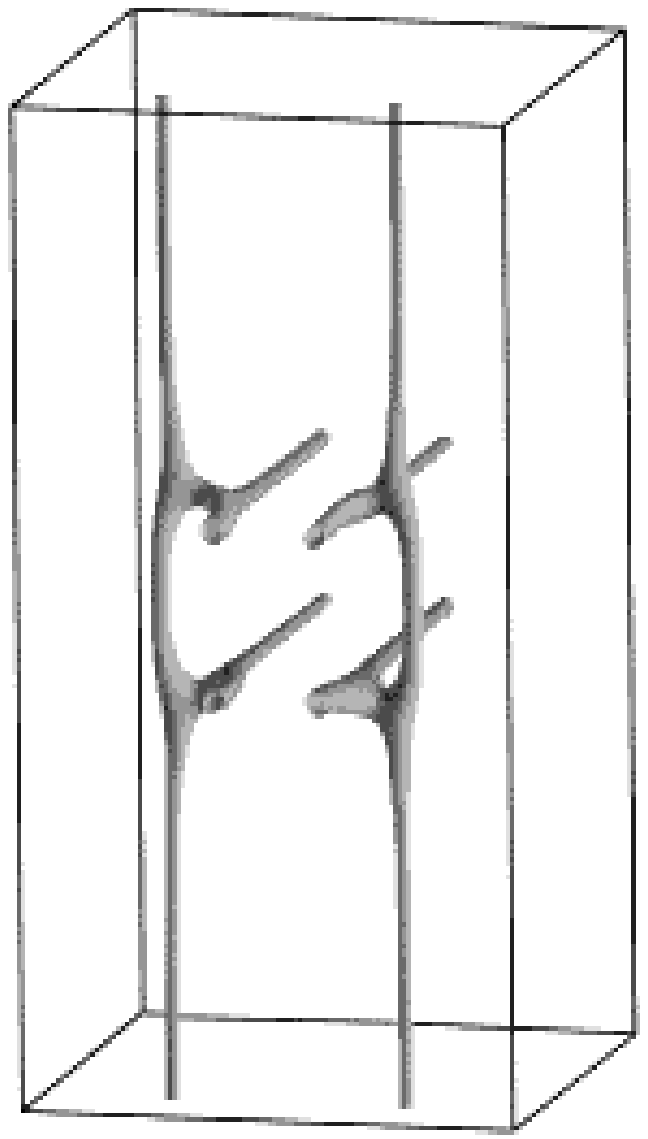}
    (c)
   \end{center}
  \end{minipage}
 \end{center}
 \begin{center}
  \begin{minipage}{0.47\linewidth}
   \begin{center}
    \epsfxsize=\linewidth \epsfbox{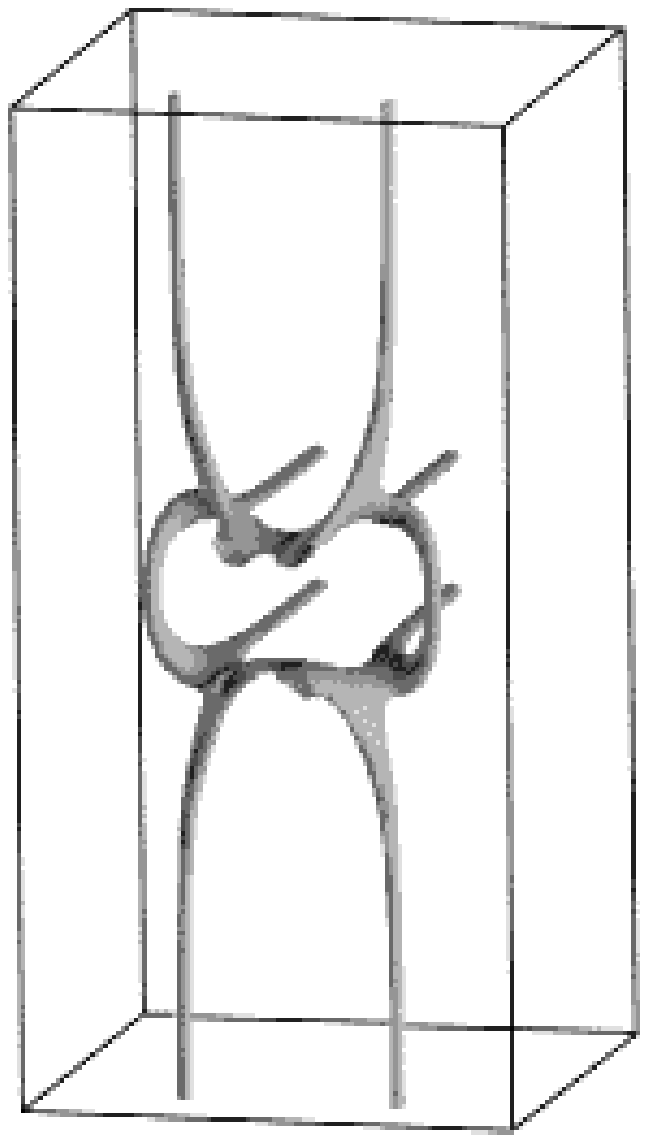}
    (d)
   \end{center}
  \end{minipage}
  \begin{minipage}{0.47\linewidth}
   \begin{center}
    \epsfxsize=\linewidth \epsfbox{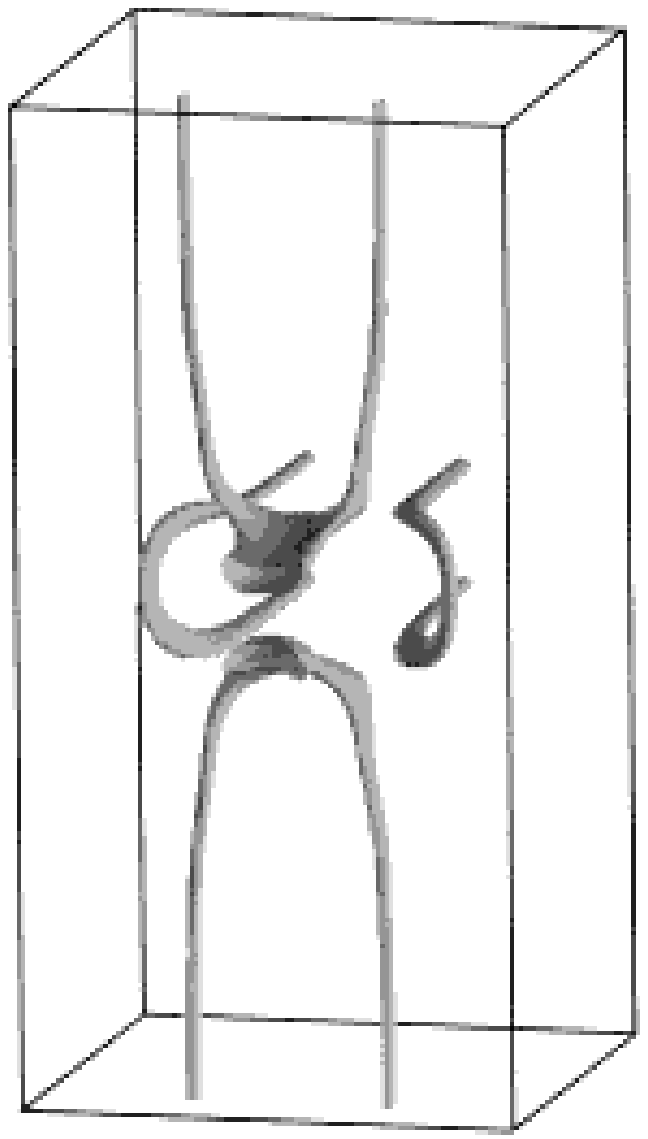}
    (e)
   \end{center}
  \end{minipage}
 \end{center}
 \begin{center}
  \begin{minipage}{0.47\linewidth}
   \begin{center}
    \epsfxsize=\linewidth \epsfbox{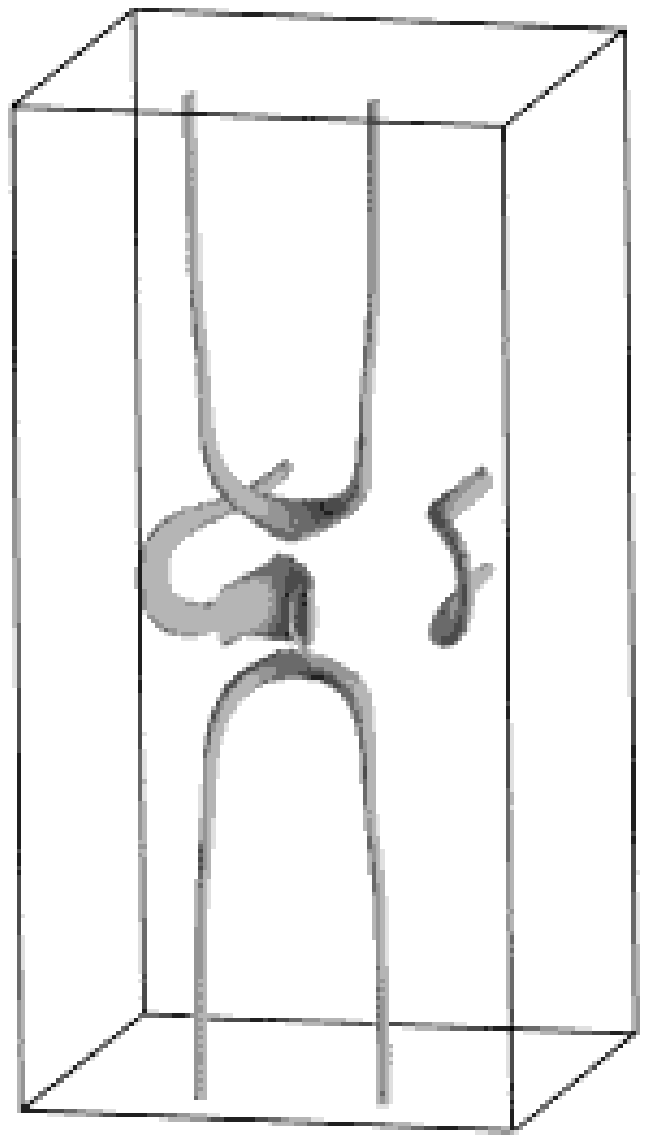}
    (f)
   \end{center}
  \end{minipage}
  \begin{minipage}{0.47\linewidth}
   \begin{center}
    \epsfxsize=\linewidth \epsfbox{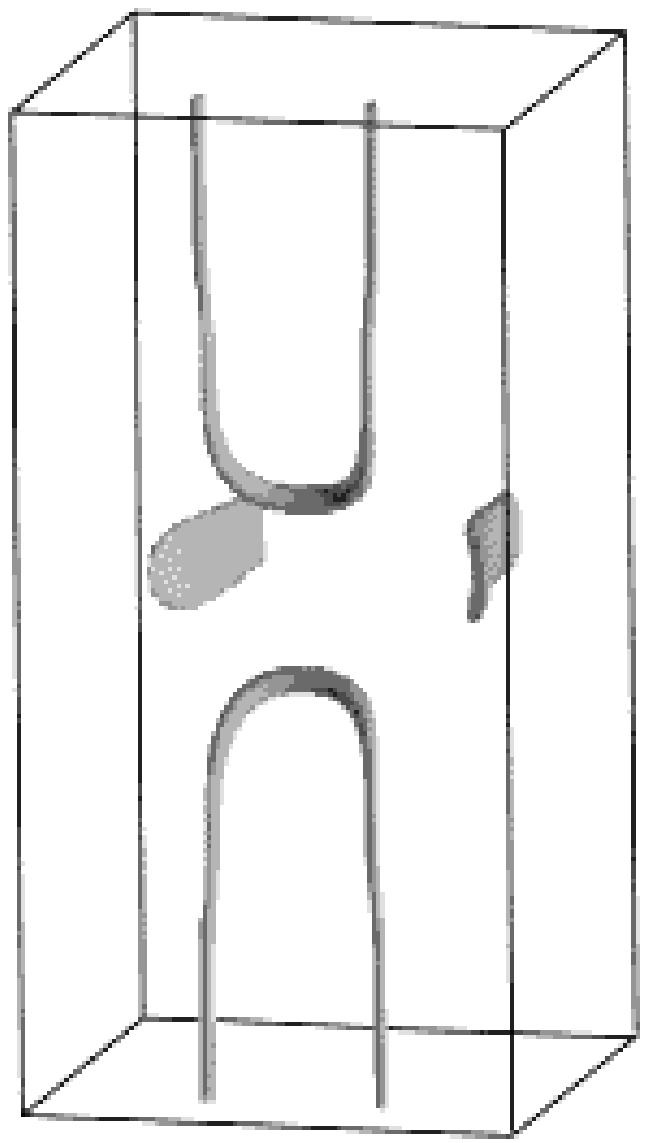}
    (g)
   \end{center}
  \end{minipage}
 \end{center}
 \begin{center}
  \begin{minipage}{0.47\linewidth}
   \begin{center}
    \epsfxsize=\linewidth \epsfbox{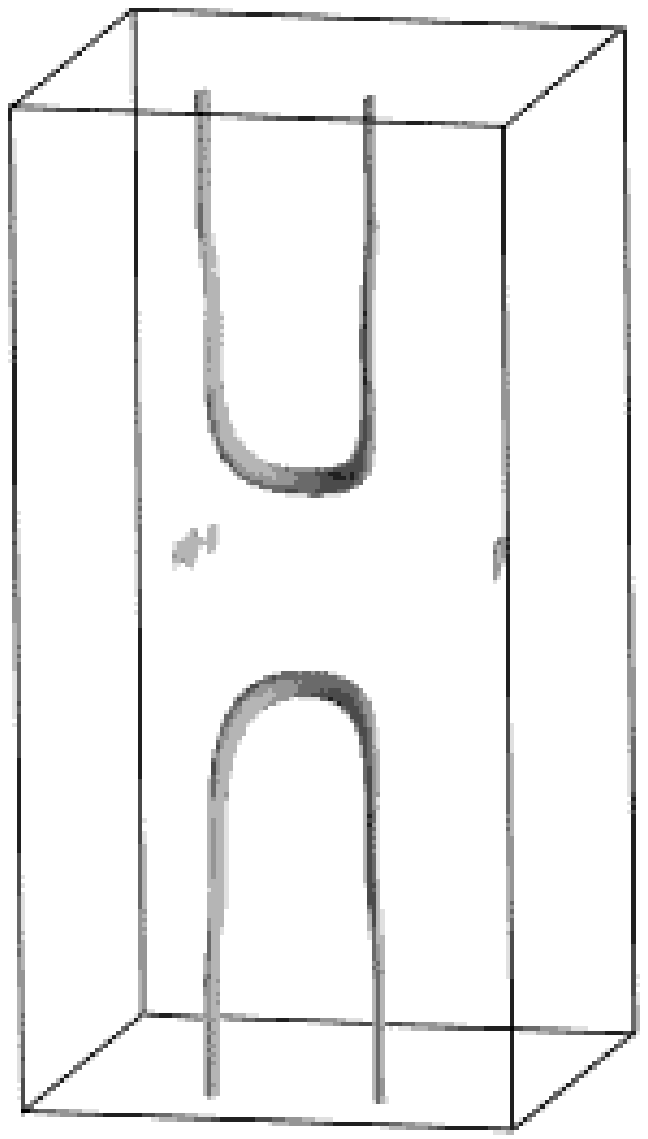}
    (h)
   \end{center}
  \end{minipage}
  \begin{minipage}{0.47\linewidth}
   \begin{center}
    \epsfxsize=\linewidth \epsfbox{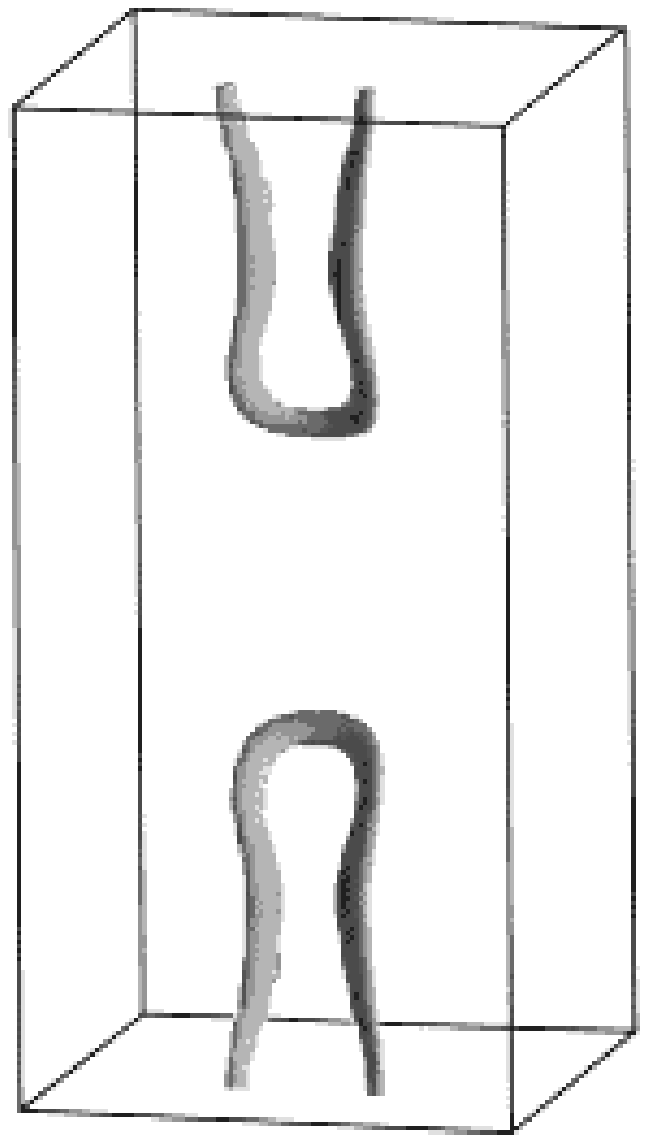}
    (i)
   \end{center}
  \end{minipage}
 \end{center}
 \begin{center}
  \begin{minipage}{0.47\linewidth}
   \begin{center}
    \epsfxsize=\linewidth \epsfbox{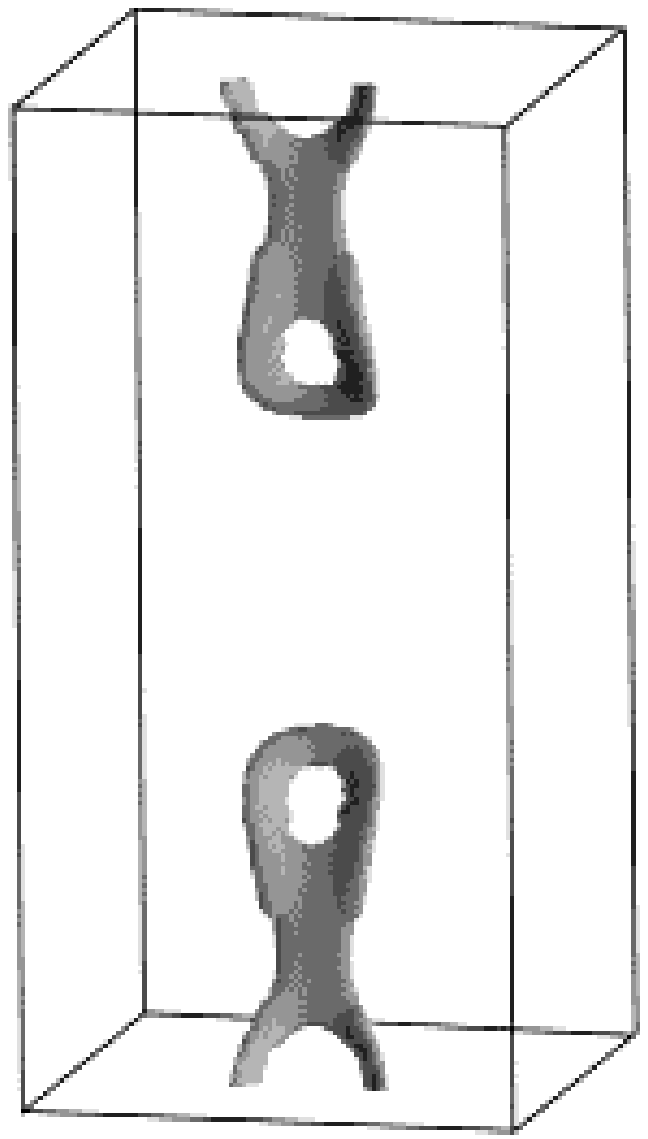}
    (j)
   \end{center}
  \end{minipage}
  \begin{minipage}{0.47\linewidth}
   \begin{center}
    \epsfxsize=\linewidth \epsfbox{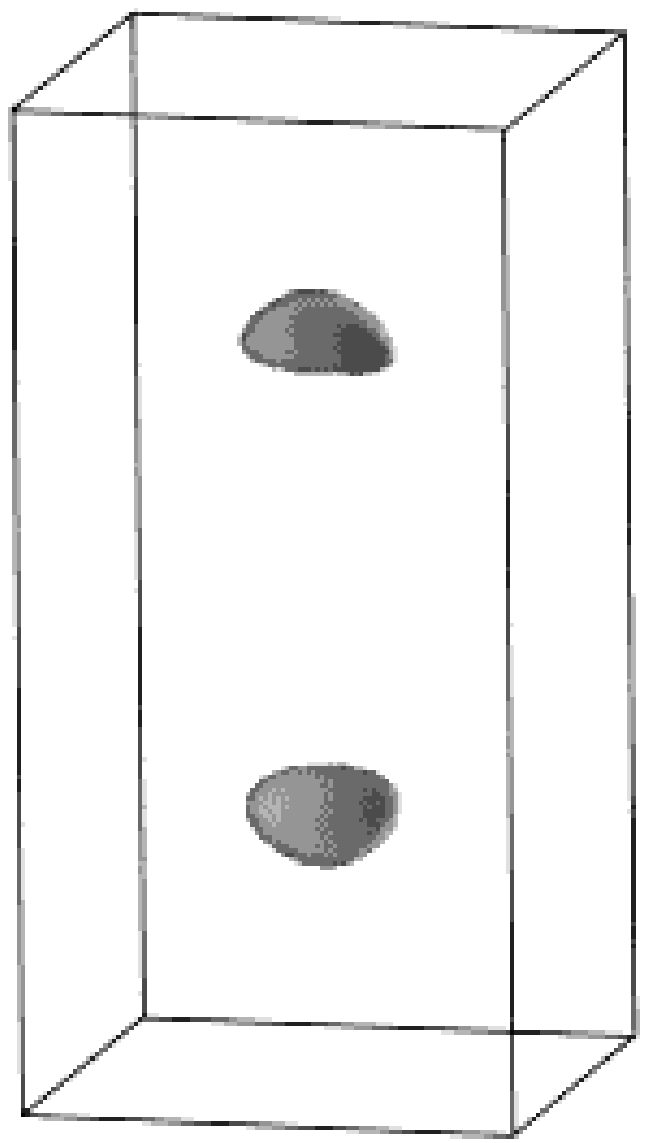}
    (k)
   \end{center}
  \end{minipage}
 \end{center}
 \caption{Reconnections of two vortices, $t=0.0$(a,b), $0.182$(c),
 $0.58$(d), $0.928$(e), $1.326$(f), $1.794$(g), $2.132$(h),
 $3.498$(i), $3.948$(j), $5.172$(k). The surface represents the contour
 of $|f|=0.06$. The inside box of (a) indicates the region plotted in
 (b)-(k).}
 \label{f11}
\end{figure}

Since the sound wave propagates over the density profile of the vortex
and the vortex itself moves, we may find more easily the sound wave
propagation in the region where the density of the condensate is nearly
homogeneous.
Thus in order to find the sound wave propagation after the
reconnection of 90$^{\circ}$ vortices under the periodic boundary
condition, we shift the reconnecting point toward the corner of the
periodic box in this section.
The initial configuration is shown in Fig. \ref{f11}(a).
The vortices are placed at a right angle, one has the singular core
along $-\hat{z}$ direction at $(x,y)=(2.5,1.9)$, other has that
along $\hat{y}$ direction at $(x,z)=(1.3,1.9)$.
In this configuration, we may show the density propagation enough far form
the singular core.

Figure \ref{f11} shows the dynamics of two very close vortices
meeting at a right angle.
These figures show that the vortices are reduced to the vortex rings
through the reconnections and disappear eventually.
This processes is the cascade process proposed by Feynman\cite{Feynman}
and revealed recently by Tsubota {\it et. al}\cite{Tsubota}.
For example, the vortices are broken up to
smaller loops by the reconnection((d)-(e)).
A small one makes self-reconnection and disappears((f)-(g)),
when the energy of vortices is transferred into the kinetic energy.
This is just the final stage of the cascade process.
At very low temperatures, these mechanism should play a crucial role in
the decay of the vortex tangle.

\begin{figure}[p]
\begin{center}
 \epsfxsize=\linewidth \epsfbox{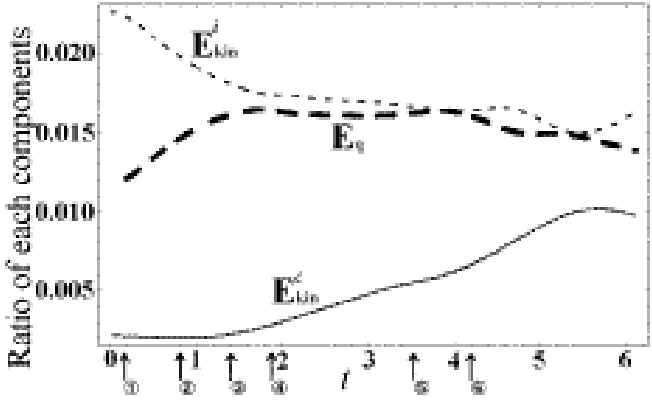}
\end{center}
 \caption{Changes in each energy component of Fig. \ref{f11}. The thick
 broken line ,the thin broken line and the 
 dotted line show respectively
 $E_{\rm q}/E_{\rm tot},E_{\rm kin}^i/E_{\rm tot}$
 and $E_{\rm kin}^i/E_{\rm tot}$. The
 numerics on the bottom show when the reconnection occurs.}
 \label{f05}
\end{figure}

The changes in the ratio of each energy component are
shown in Fig. \ref{f05}. The figures below the horizontal axis
indicate when the reconnections occur.

\begin{figure}[p]
 \begin{center}
  \begin{minipage}{0.47\linewidth}
   \begin{center}
    \epsfxsize=\linewidth \epsfbox{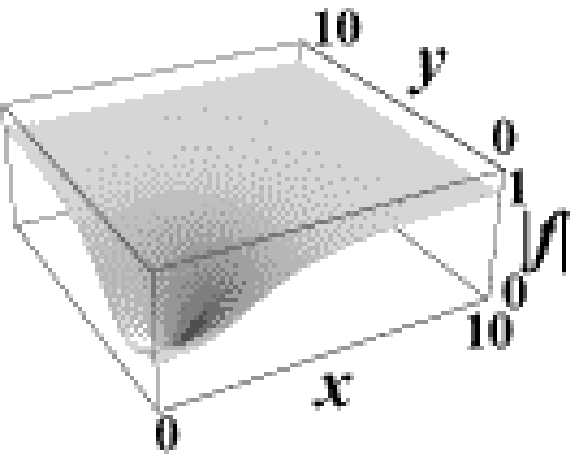}
    (a)
   \end{center}
  \end{minipage}
  \begin{minipage}{0.47\linewidth}
   \begin{center}
    \epsfxsize=\linewidth \epsfbox{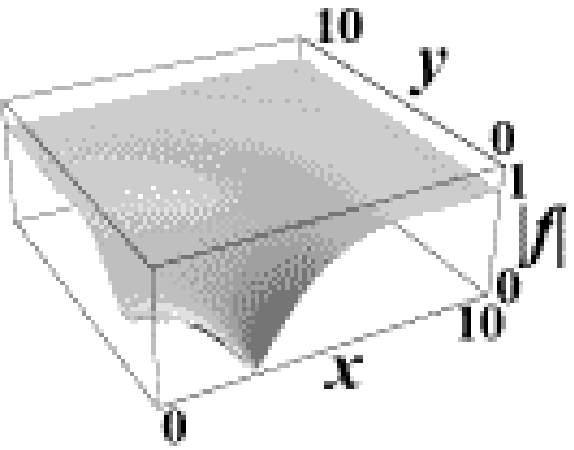}
    (b)
   \end{center}
  \end{minipage}
 \end{center}
 \begin{center}
  \begin{minipage}{0.47\linewidth}
   \begin{center}
    \epsfxsize=\linewidth \epsfbox{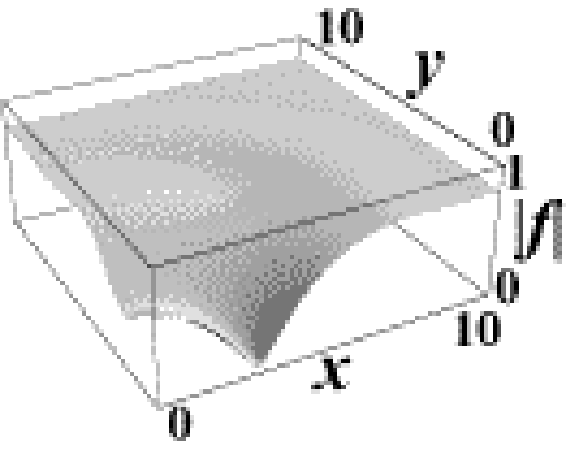}
    (c)
   \end{center}
  \end{minipage}
  \begin{minipage}{0.47\linewidth}
   \begin{center}
    \epsfxsize=\linewidth \epsfbox{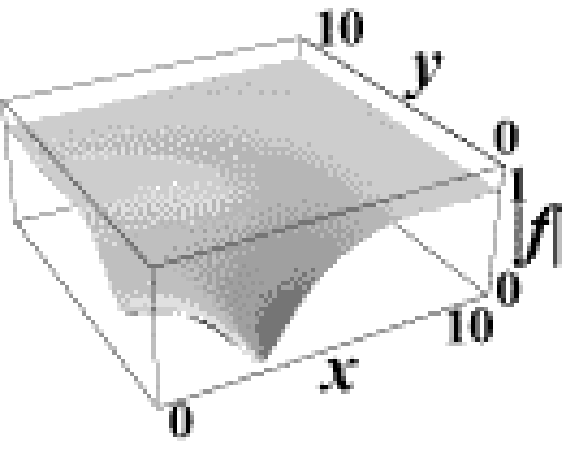}
    (d)
   \end{center}
  \end{minipage}
 \end{center}
 \caption{Profiles of $|f|$. The vertical axis refers to $|f|$, and the
 others represents the space coordinates, , $t=0.0$(a), $1.938$(b),
 $2.04$(c), $2.142$(d). The dimly seeable wakes are sound waves((b)-(d)).}
 \label{f12}
\end{figure}

\begin{figure}[p]
 \begin{center}
  \begin{minipage}{0.47\linewidth}
   \begin{center}
    \epsfxsize=\linewidth \epsfbox{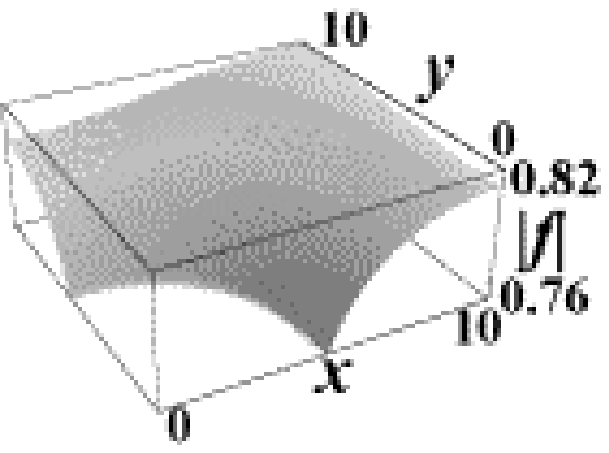}
    (a)
   \end{center}
  \end{minipage}
  \begin{minipage}{0.47\linewidth}
   \begin{center}
    \epsfxsize=\linewidth \epsfbox{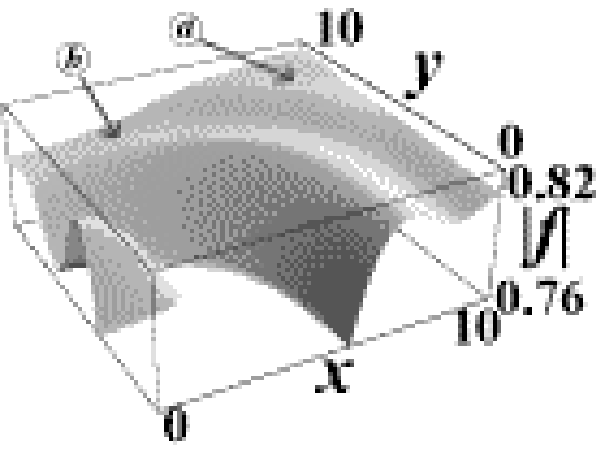}
    (b)
   \end{center}
  \end{minipage}
 \end{center}
 \begin{center}
  \begin{minipage}{0.47\linewidth}
   \begin{center}
    \epsfxsize=\linewidth \epsfbox{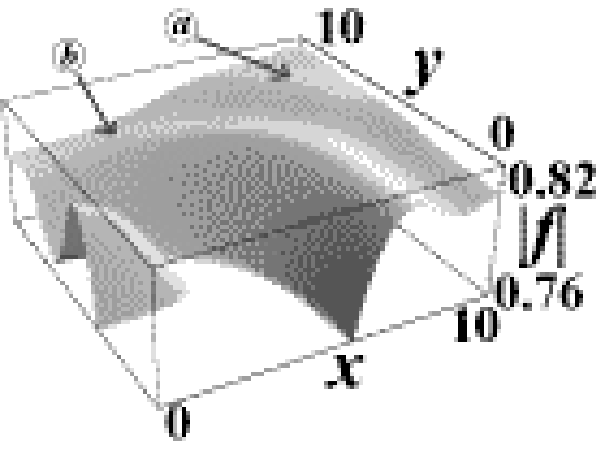}
    (c)
   \end{center}
  \end{minipage}
  \begin{minipage}{0.47\linewidth}
   \begin{center}
    \epsfxsize=\linewidth \epsfbox{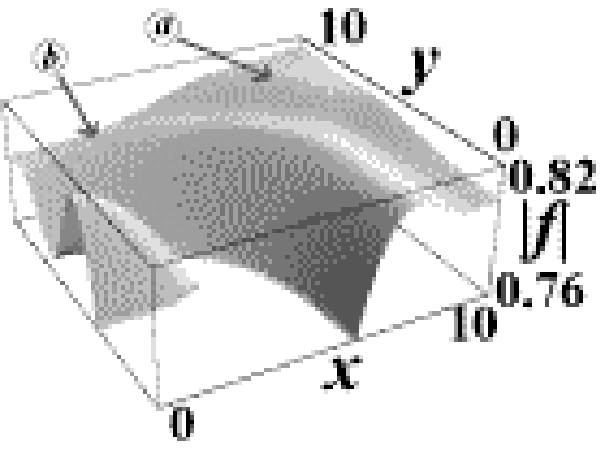}
    (d)
   \end{center}
  \end{minipage}
 \end{center}
 \caption{Sectional profile of $|f|$. The range of $|f|$ is limited to
 $0.76\leq|f|\leq0.82$. Coordinates are the same as
 Fig. \ref{f12}, $t=0.0$(a), $1.938$(b), $2.04$(c), $2.142$(d). Some
 propagating wakes are sound waves.}
 \label{f13}
\end{figure}

The density propagation is visible enough far from the vortex core.
The dimly seeable wakes are shown in Fig. \ref{f12}.
In Fig. \ref{f13} which enlarges the sectional
profiles of Fig. \ref{f12}, the propagation of the wakes
are shown properly.
From the analysis of the speed of the sound wave in Sec. \ref{sound},
we find that the sound wave \circlenum{a} and \circlenum{b}
probably comes from
the reconnection \circlenum{2} and \circlenum{3}, respectively.
As described before, the amplitude of the sound wave depends on
how much the quantum energy is transferred into the kinetic energy,
in other words,
the vortex line length is lost by the reconnection.
The reconnection \circlenum{2} is the point reconnection, while
\circlenum{3} is the annihilation of a small vortex ring.
Although the lost line length of each two reconnection is shorter than
that in Sec. \ref{sound}, the sound wave propagations are obviously
visible. This is because the density enough far from the singular cores
is about homogeneous.
For example in $5\leq x\leq 10$ and $5\leq y\leq 10$, $|f|\sim 0.8$.
%Thus in order to show the sound wave propagation,
%it is necessary to consider the proper initial configuration of
%the singular vortex cores.

\section{Change in the energy components}\label{energy}
The comparison between Fig. \ref{f09}, Fig. \ref{f04} and Fig. \ref{f05}
shows some important properties on how the energy components behave. The
energy is generally transferred between these three components. In fact,
the changes in the components after the reconnection are roughly
classified into two stages. In the first stage, $E_{\rm q}$ increases,
as shown in Fig. \ref{f04}, and the period $0\leq t \leq 1.2$ of
Fig. \ref{f05}. Then, the vortices reconnect at points and still survive
after that. The second stage is characterized by the reduction of
$E_{\rm q}$ in the period $0.2\leq t$ of Fig. \ref{f09}, and the period
$1.2\leq t$ of Fig. \ref{f05}: the vortex cores rather collide and
disappear, than reconnect at points.

The behavior of the energy components in the first stage may be
attributed at the following interaction between the vortices and the
sound waves. While two vortices reconnect at a point, as shown in
Fig. \ref{f01}, they stretch and fat near the reconnection point because
of the inter-vortex interaction.
The vortices near the reconnection point move with the velocity
comparable to the sound velocity. The sound waves made by the
reconnection, of course, expand outside, but the above factors make the
sound waves interact continuously with the vortex cores even after the
reconnection. Although we do not know the detail of the interaction
between quantized vortices and sound waves, this continuous interaction
must make it easy that the sound waves are absorbed by the vortex cores,
which increases $E_{\rm q}$.

On the other hand, the vortex cores fade out in the second stage. Thus
the quantum energy decreases, and the kinetic energy increase owing to
the acoustic emission. However, $\nabla \rho$ never vanishes even after
the core structure disappears, so that the energy components are mixed
and oscillate.

\section{Kelvin's circulation theorem}\label{kel}
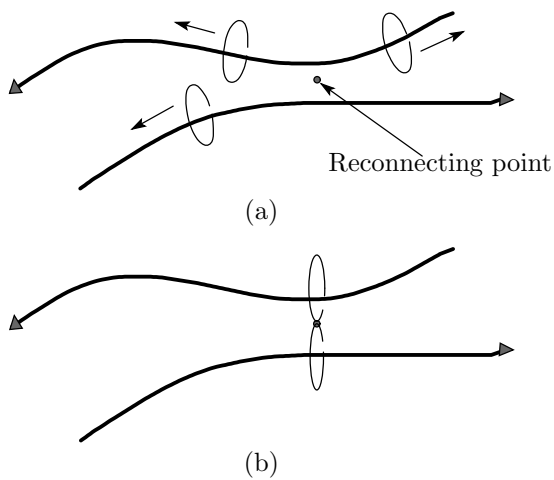
\begin{figure}[p]
 \begin{center}
  %WinTpicVersion2.15
\unitlength 0.1in
\begin{picture}(26.37,9.28)(3.30,-9.90)
% SPLINE 0 0 3 0
% 5 341 880 863 604 2009 730 2621 490 2651 472
% 
\special{pn 20}%
\special{pa 341 480}%
\special{pa 369 461}%
\special{pa 396 441}%
\special{pa 424 422}%
\special{pa 451 403}%
\special{pa 479 385}%
\special{pa 507 367}%
\special{pa 535 349}%
\special{pa 563 332}%
\special{pa 591 315}%
\special{pa 619 299}%
\special{pa 647 284}%
\special{pa 675 270}%
\special{pa 704 257}%
\special{pa 733 244}%
\special{pa 762 233}%
\special{pa 791 223}%
\special{pa 820 214}%
\special{pa 850 207}%
\special{pa 879 201}%
\special{pa 909 196}%
\special{pa 940 193}%
\special{pa 970 191}%
\special{pa 1001 190}%
\special{pa 1031 190}%
\special{pa 1062 191}%
\special{pa 1094 193}%
\special{pa 1125 196}%
\special{pa 1156 199}%
\special{pa 1188 204}%
\special{pa 1220 209}%
\special{pa 1252 215}%
\special{pa 1284 221}%
\special{pa 1316 227}%
\special{pa 1348 234}%
\special{pa 1380 241}%
\special{pa 1413 249}%
\special{pa 1445 256}%
\special{pa 1477 264}%
\special{pa 1510 271}%
\special{pa 1542 279}%
\special{pa 1575 286}%
\special{pa 1607 293}%
\special{pa 1640 300}%
\special{pa 1672 306}%
\special{pa 1705 312}%
\special{pa 1737 317}%
\special{pa 1769 322}%
\special{pa 1802 326}%
\special{pa 1834 329}%
\special{pa 1866 332}%
\special{pa 1898 333}%
\special{pa 1930 334}%
\special{pa 1962 333}%
\special{pa 1994 331}%
\special{pa 2025 328}%
\special{pa 2057 324}%
\special{pa 2088 319}%
\special{pa 2119 312}%
\special{pa 2150 305}%
\special{pa 2181 297}%
\special{pa 2212 287}%
\special{pa 2242 277}%
\special{pa 2273 266}%
\special{pa 2303 254}%
\special{pa 2333 242}%
\special{pa 2363 228}%
\special{pa 2392 215}%
\special{pa 2422 200}%
\special{pa 2451 186}%
\special{pa 2480 170}%
\special{pa 2508 155}%
\special{pa 2537 139}%
\special{pa 2565 123}%
\special{pa 2593 107}%
\special{pa 2620 90}%
\special{pa 2648 74}%
\special{pa 2651 72}%
\special{sp}%
% SPLINE 0 0 3 0
% 4 700 1390 1366 1000 2770 916 2926 916
% 
\special{pn 20}%
\special{pa 700 990}%
\special{pa 727 971}%
\special{pa 753 952}%
\special{pa 780 933}%
\special{pa 807 915}%
\special{pa 834 896}%
\special{pa 861 877}%
\special{pa 888 859}%
\special{pa 915 841}%
\special{pa 942 823}%
\special{pa 969 805}%
\special{pa 996 788}%
\special{pa 1023 771}%
\special{pa 1051 754}%
\special{pa 1078 738}%
\special{pa 1106 722}%
\special{pa 1134 706}%
\special{pa 1162 691}%
\special{pa 1190 677}%
\special{pa 1218 663}%
\special{pa 1247 649}%
\special{pa 1275 636}%
\special{pa 1304 624}%
\special{pa 1333 612}%
\special{pa 1363 601}%
\special{pa 1392 591}%
\special{pa 1422 581}%
\special{pa 1452 572}%
\special{pa 1482 564}%
\special{pa 1512 556}%
\special{pa 1542 549}%
\special{pa 1573 542}%
\special{pa 1604 536}%
\special{pa 1635 531}%
\special{pa 1666 526}%
\special{pa 1697 521}%
\special{pa 1729 517}%
\special{pa 1761 513}%
\special{pa 1792 510}%
\special{pa 1824 507}%
\special{pa 1856 505}%
\special{pa 1888 503}%
\special{pa 1920 501}%
\special{pa 1953 500}%
\special{pa 1985 499}%
\special{pa 2017 498}%
\special{pa 2050 498}%
\special{pa 2082 498}%
\special{pa 2115 498}%
\special{pa 2148 498}%
\special{pa 2180 499}%
\special{pa 2213 499}%
\special{pa 2246 500}%
\special{pa 2279 501}%
\special{pa 2311 502}%
\special{pa 2344 503}%
\special{pa 2377 504}%
\special{pa 2410 505}%
\special{pa 2442 506}%
\special{pa 2475 508}%
\special{pa 2508 509}%
\special{pa 2541 510}%
\special{pa 2573 511}%
\special{pa 2606 512}%
\special{pa 2638 513}%
\special{pa 2671 514}%
\special{pa 2703 515}%
\special{pa 2735 516}%
\special{pa 2768 516}%
\special{pa 2800 516}%
\special{pa 2832 516}%
\special{pa 2864 516}%
\special{pa 2896 516}%
\special{pa 2926 516}%
\special{sp}%
% CIRCLE 2 0 0 0
% 4 1940 820 1958 820 1958 820 1958 820
% 
\special{pn 8}%
\special{sh 0.600}%
\special{ar 1940 420 18 18  0.0000000 6.2831853}%
% VECTOR 2 0 3 0
% 2 2490 1220 1962 842
% 
\special{pn 8}%
\special{pa 2490 820}%
\special{pa 1962 442}%
\special{fp}%
\special{sh 1}%
\special{pa 1962 442}%
\special{pa 2005 497}%
\special{pa 2005 473}%
\special{pa 2028 465}%
\special{pa 1962 442}%
\special{fp}%
% STR 2 0 3 0
% 3 2300 1240 2300 1300 2 0
% Reconnection point
\put(20.0000,-9.0000){\makebox(0,0)[lb]{Reconnecting point}}%
% SPLINE 2 0 3 0
% 42 2424 623 2430 648 2435 673 2438 697 2439 719 2438 739 2435 756 2431 771 2425 782 2417 790 2408 795 2398 796 2387 793 2375 786 2363 776 2351 762 2338 746 2327 726 2316 704 2305 681 2296 657 2289 632 2282 607 2278 582 2275 559 2275 537 2276 517 2278 500 2284 485 2289 474 2297 467 2306 463 2317 463 2327 466 2339 473 2351 484 2364 497 2377 514 2387 534 2399 556 2409 580 2409 580
% 
\special{pn 8}%
\special{pa 2424 223}%
\special{pa 2431 254}%
\special{pa 2437 286}%
\special{pa 2439 318}%
\special{pa 2436 349}%
\special{pa 2427 380}%
\special{pa 2400 396}%
\special{pa 2372 383}%
\special{pa 2350 360}%
\special{pa 2331 334}%
\special{pa 2317 306}%
\special{pa 2303 277}%
\special{pa 2293 246}%
\special{pa 2284 216}%
\special{pa 2278 184}%
\special{pa 2275 152}%
\special{pa 2276 120}%
\special{pa 2282 89}%
\special{pa 2301 65}%
\special{pa 2332 68}%
\special{pa 2356 89}%
\special{pa 2377 113}%
\special{pa 2391 142}%
\special{pa 2405 171}%
\special{pa 2409 180}%
\special{sp}%
% SPLINE 2 0 3 0
% 42 1390 1000 1395 1026 1399 1051 1401 1074 1401 1096 1400 1116 1396 1134 1391 1149 1385 1159 1377 1167 1367 1172 1358 1172 1346 1168 1335 1161 1323 1150 1311 1136 1299 1119 1289 1099 1278 1077 1269 1054 1261 1029 1255 1004 1250 979 1246 954 1244 930 1244 908 1247 888 1250 871 1256 857 1262 846 1270 839 1279 835 1289 835 1300 839 1312 847 1323 858 1335 873 1347 890 1358 910 1368 932 1377 957 1377 957
% 
\special{pn 8}%
\special{pa 1390 600}%
\special{pa 1396 631}%
\special{pa 1400 663}%
\special{pa 1401 695}%
\special{pa 1398 727}%
\special{pa 1387 757}%
\special{pa 1360 772}%
\special{pa 1332 758}%
\special{pa 1310 735}%
\special{pa 1293 708}%
\special{pa 1279 679}%
\special{pa 1267 649}%
\special{pa 1258 619}%
\special{pa 1252 587}%
\special{pa 1246 556}%
\special{pa 1244 524}%
\special{pa 1246 492}%
\special{pa 1254 461}%
\special{pa 1274 437}%
\special{pa 1304 441}%
\special{pa 1327 463}%
\special{pa 1346 489}%
\special{pa 1361 517}%
\special{pa 1373 547}%
\special{pa 1377 557}%
\special{sp}%
% SPLINE 2 0 3 0
% 42 1572 695 1567 720 1561 745 1553 767 1544 788 1535 806 1525 821 1514 832 1504 839 1494 844 1484 844 1475 840 1466 832 1458 821 1451 807 1446 789 1442 769 1439 746 1439 722 1439 697 1442 671 1446 646 1451 620 1458 596 1465 574 1474 553 1483 536 1493 521 1504 511 1514 503 1524 499 1534 500 1543 504 1552 512 1559 524 1566 538 1571 556 1576 577 1577 599 1578 623 1577 650 1577 650
% 
\special{pn 8}%
\special{pa 1572 295}%
\special{pa 1566 326}%
\special{pa 1557 357}%
\special{pa 1545 387}%
\special{pa 1530 415}%
\special{pa 1507 437}%
\special{pa 1477 441}%
\special{pa 1456 418}%
\special{pa 1446 387}%
\special{pa 1440 356}%
\special{pa 1439 324}%
\special{pa 1439 292}%
\special{pa 1444 260}%
\special{pa 1449 229}%
\special{pa 1457 198}%
\special{pa 1467 167}%
\special{pa 1481 139}%
\special{pa 1501 114}%
\special{pa 1528 99}%
\special{pa 1555 116}%
\special{pa 1568 145}%
\special{pa 1576 176}%
\special{pa 1577 208}%
\special{pa 1578 240}%
\special{pa 1577 250}%
\special{sp}%
% VECTOR 2 0 3 0
% 6 1404 575 1200 521 1188 959 972 1079 2484 671 2706 569
% 
\special{pn 8}%
\special{pa 1404 175}%
\special{pa 1200 121}%
\special{fp}%
\special{sh 1}%
\special{pa 1200 121}%
\special{pa 1259 157}%
\special{pa 1252 135}%
\special{pa 1270 119}%
\special{pa 1200 121}%
\special{fp}%
\special{pa 1188 559}%
\special{pa 972 679}%
\special{fp}%
\special{sh 1}%
\special{pa 972 679}%
\special{pa 1040 664}%
\special{pa 1019 653}%
\special{pa 1021 629}%
\special{pa 972 679}%
\special{fp}%
\special{pa 2484 271}%
\special{pa 2706 169}%
\special{fp}%
\special{sh 1}%
\special{pa 2706 169}%
\special{pa 2637 179}%
\special{pa 2658 191}%
\special{pa 2654 215}%
\special{pa 2706 169}%
\special{fp}%
% POLYGON 2 0 0 0
% 7 330 893 354 821 396 887 396 887 330 893 330 893 330 893
% 
\special{pn 8}%
\special{sh 0.600}%
\special{pa 330 493}%
\special{pa 354 421}%
\special{pa 396 487}%
\special{pa 396 487}%
\special{pa 330 493}%
\special{pa 330 493}%
\special{pa 330 493}%
\special{fp}%
% POLYGON 2 0 0 0
% 7 2900 890 2967 925 2895 956 2895 956 2900 890 2900 890 2900 890
% 
\special{pn 8}%
\special{sh 0.600}%
\special{pa 2900 490}%
\special{pa 2967 525}%
\special{pa 2895 556}%
\special{pa 2895 556}%
\special{pa 2900 490}%
\special{pa 2900 490}%
\special{pa 2900 490}%
\special{fp}%
\end{picture}%

  (a)
 \end{center}
 \begin{center}
  %WinTpicVersion2.15
\unitlength 0.1in
\begin{picture}(26.35,10.03)(3.30,-10.82)
% SPLINE 0 0 3 0
% 5 341 887 863 611 2009 737 2621 497 2651 479
% 
\special{pn 20}%
\special{pa 341 487}%
\special{pa 369 468}%
\special{pa 396 448}%
\special{pa 424 429}%
\special{pa 451 410}%
\special{pa 479 392}%
\special{pa 507 374}%
\special{pa 535 356}%
\special{pa 563 339}%
\special{pa 591 322}%
\special{pa 619 306}%
\special{pa 647 291}%
\special{pa 675 277}%
\special{pa 704 264}%
\special{pa 733 251}%
\special{pa 762 240}%
\special{pa 791 230}%
\special{pa 820 221}%
\special{pa 850 214}%
\special{pa 879 208}%
\special{pa 909 203}%
\special{pa 940 200}%
\special{pa 970 198}%
\special{pa 1001 197}%
\special{pa 1031 197}%
\special{pa 1062 198}%
\special{pa 1094 200}%
\special{pa 1125 203}%
\special{pa 1156 206}%
\special{pa 1188 211}%
\special{pa 1220 216}%
\special{pa 1252 222}%
\special{pa 1284 228}%
\special{pa 1316 234}%
\special{pa 1348 241}%
\special{pa 1380 248}%
\special{pa 1413 256}%
\special{pa 1445 263}%
\special{pa 1477 271}%
\special{pa 1510 278}%
\special{pa 1542 286}%
\special{pa 1575 293}%
\special{pa 1607 300}%
\special{pa 1640 307}%
\special{pa 1672 313}%
\special{pa 1705 319}%
\special{pa 1737 324}%
\special{pa 1769 329}%
\special{pa 1802 333}%
\special{pa 1834 336}%
\special{pa 1866 339}%
\special{pa 1898 340}%
\special{pa 1930 341}%
\special{pa 1962 340}%
\special{pa 1994 338}%
\special{pa 2025 335}%
\special{pa 2057 331}%
\special{pa 2088 326}%
\special{pa 2119 319}%
\special{pa 2150 312}%
\special{pa 2181 304}%
\special{pa 2212 294}%
\special{pa 2242 284}%
\special{pa 2273 273}%
\special{pa 2303 261}%
\special{pa 2333 249}%
\special{pa 2363 235}%
\special{pa 2392 222}%
\special{pa 2422 207}%
\special{pa 2451 193}%
\special{pa 2480 177}%
\special{pa 2508 162}%
\special{pa 2537 146}%
\special{pa 2565 130}%
\special{pa 2593 114}%
\special{pa 2620 97}%
\special{pa 2648 81}%
\special{pa 2651 79}%
\special{sp}%
% SPLINE 0 0 3 0
% 4 702 1482 1368 1092 2772 1008 2928 1008
% 
\special{pn 20}%
\special{pa 702 1082}%
\special{pa 729 1063}%
\special{pa 755 1044}%
\special{pa 782 1025}%
\special{pa 809 1007}%
\special{pa 836 988}%
\special{pa 863 969}%
\special{pa 890 951}%
\special{pa 917 933}%
\special{pa 944 915}%
\special{pa 971 897}%
\special{pa 998 880}%
\special{pa 1025 863}%
\special{pa 1053 846}%
\special{pa 1080 830}%
\special{pa 1108 814}%
\special{pa 1136 798}%
\special{pa 1164 783}%
\special{pa 1192 769}%
\special{pa 1220 755}%
\special{pa 1249 741}%
\special{pa 1277 728}%
\special{pa 1306 716}%
\special{pa 1335 704}%
\special{pa 1365 693}%
\special{pa 1394 683}%
\special{pa 1424 673}%
\special{pa 1454 664}%
\special{pa 1484 656}%
\special{pa 1514 648}%
\special{pa 1544 641}%
\special{pa 1575 634}%
\special{pa 1606 628}%
\special{pa 1637 623}%
\special{pa 1668 618}%
\special{pa 1699 613}%
\special{pa 1731 609}%
\special{pa 1763 605}%
\special{pa 1794 602}%
\special{pa 1826 599}%
\special{pa 1858 597}%
\special{pa 1890 595}%
\special{pa 1922 593}%
\special{pa 1955 592}%
\special{pa 1987 591}%
\special{pa 2019 590}%
\special{pa 2052 590}%
\special{pa 2084 590}%
\special{pa 2117 590}%
\special{pa 2150 590}%
\special{pa 2182 591}%
\special{pa 2215 591}%
\special{pa 2248 592}%
\special{pa 2281 593}%
\special{pa 2313 594}%
\special{pa 2346 595}%
\special{pa 2379 596}%
\special{pa 2412 597}%
\special{pa 2444 598}%
\special{pa 2477 600}%
\special{pa 2510 601}%
\special{pa 2543 602}%
\special{pa 2575 603}%
\special{pa 2608 604}%
\special{pa 2640 605}%
\special{pa 2673 606}%
\special{pa 2705 607}%
\special{pa 2737 608}%
\special{pa 2770 608}%
\special{pa 2802 608}%
\special{pa 2834 608}%
\special{pa 2866 608}%
\special{pa 2898 608}%
\special{pa 2928 608}%
\special{sp}%
% CIRCLE 2 0 0 0
% 4 1938 870 1956 870 1956 870 1956 870
% 
\special{pn 8}%
\special{sh 0.600}%
\special{ar 1938 470 18 18  0.0000000 6.2831853}%
% POLYGON 2 0 0 0
% 7 330 900 354 828 396 894 396 894 330 900 330 900 330 900
% 
\special{pn 8}%
\special{sh 0.600}%
\special{pa 330 500}%
\special{pa 354 428}%
\special{pa 396 494}%
\special{pa 396 494}%
\special{pa 330 500}%
\special{pa 330 500}%
\special{pa 330 500}%
\special{fp}%
% POLYGON 2 0 0 0
% 7 2898 972 2965 1007 2893 1038 2893 1038 2898 972 2898 972 2898 972
% 
\special{pn 8}%
\special{sh 0.600}%
\special{pa 2898 572}%
\special{pa 2965 607}%
\special{pa 2893 638}%
\special{pa 2893 638}%
\special{pa 2898 572}%
\special{pa 2898 572}%
\special{pa 2898 572}%
\special{fp}%
% ELLIPSE 2 0 3 0
% 4 1938 1044 1974 1218 2046 762 2184 870
% 
\special{pn 8}%
\special{ar 1938 644 36 174  6.1378753 6.2831853}%
\special{ar 1938 644 36 174  0.0000000 5.7903489}%
% ELLIPSE 2 0 3 0
% 4 1938 684 1980 858 2046 792 2004 924
% 
\special{pn 8}%
\special{ar 1938 284 42 174  0.7209713 6.2831853}%
\special{ar 1938 284 42 174  0.0000000 0.2362453}%
\end{picture}%

  (b)
 \end{center}
 \caption{Motion of the material loops when two vortices
 reconnect. (a):most loops are carried away from the merger region by
 the flow. (b):the Kelvin's circulation theorem breaks on the loops going
 through the reconnecting point.}
 \label{f03}
\end{figure}

This section comments on the conservation of the circulation
$\Gamma=\oint d\mbf{s}\cdot\mbf{v}$, which is called the Kelvin's
circulation theorem.
This theorem states that the circulation round any closed material loop
is invariant in an inviscid(barotropic) fluid, and vortices are
never born, die and reconnect. 
It seems that the theorem does not allow the above reconnection of
quantized vortex.
Then, we have to consider two problems:(i) is
this theorem applicable to the superfluid dynamics derived from the GPE?
(ii) what happens to the material loops when quantized vortices
reconnect?
%(i)
Equation (\ref{e14}) yields the equation of motion for
superflow:
\begin{equation}
 \dfrac{D\mbf{v}}{Dt}=\nabla\left(\dfrac{\nabla^2|f|}{|f|}\right)=\nabla w,
  \label{e15}
\end{equation}
where $D/Dt\equiv\partial/\partial t+\mbf{v}\cdot\grad$.
By differentiating $\Gamma$ with respect to $t$, we obtain
\begin{equation}
 \dfrac{D\Gamma}{Dt}=\oint d\mbf{s}\cdot\nabla\left(w+\dfrac{\mbf{v}^2}2
					     \right).
 \label{e16}
\end{equation}
Therefore $\Gamma=\text{const.}$ if $w$ is well-defined on the material
loop along which $\Gamma$ is defined((i)).
%(ii)
When two vortices are about to reconnect,
almost all material loops are carried away from the merger region by the
flow(Fig. \ref{f03}(a))((ii)). The loop just going through
the reconnecting point
breaks this theorem because $w$ is not well-defined owing to $|f|=0$,
thus leading to 
the reconnection(Fig. \ref{f03}(b)). This is the peculiar situation in
superfluid. Thus the reconnection of quantized vortices is not
contradictory to the Kelvin's circulation theorem.

\section{Conclusions and discussions}\label{conc}
We studied numerically the reconnection of quantized vortices and the
concurrent acoustic emission by the analysis of the GPE. The energy
components are analyzed for every dynamics.

When two anti-parallel vortices are close within a critical distance
which is twice the healing length, they move with the velocity
comparable to the sound velocity, emit the sound waves and disappear;
this phenomena is closely concerned with the Cherenkov resonance. The
propagation of the sound waves not only appears in the density profile
of the wave function but also affects the phase field; they destroy the
circulative velocity field that belonged to the vortices originally.

Two straight vortices meeting initially at a right angle follow the
scenario similar to the vortex filaments, thus reconnecting. The
singular lines of $|f|=0$ cross once at a point, switch their topology
instantly and leave separately. This reconnection is not contradictory
to the Kelvin's circulation theorem, because the potential of the
superflow field becomes undefined at the reconnection point.

Two close straight vortices meeting at a right angle, under the periodic
boundary condition, follow a series of reconnection, breaking up to
smaller vortex loops. Eventually these loops disappear through
self-reconnections with the acoustic emission. The process transforms
irreversibly the quantum energy of the vortex cores to the kinetic
energy. This may represent the final stage of the vortex cascade
process.

The change in energy components depends on how the vortices
reconnect. When vortices survive still after reconnections, the emitted
sound waves interact continuously with the vortex cores. Then sound
waves are absorbed by the vortex cores, which increases the quantum
energy. On the other hand, when vortex loops disappear through
reconnections, the quantum energy is transferred into the kinetic energy
with the acoustic emission.

Throughout this work, it remains unresolved how sound waves interact
with quantized vortices; they may stretch and fat the vortex cores or
excite the Kelvin waves\cite{Vinen}. More detailed studies of the
interaction are being carried out.

\section*{acknowledgement}
We are grateful to W. F. Vinen, M. E. Brachet, C. S. Adams, T. Araki and
T. Tatsuta for useful discussion.

\appendix
\section*{Motion of anti-parallel vortices by the filament
 formulation}\label{motion}
\begin{figure}[h]
 \begin{center}
  \begin{minipage}{0.47\linewidth}
   \begin{center}
    \epsfxsize=\linewidth \epsfbox{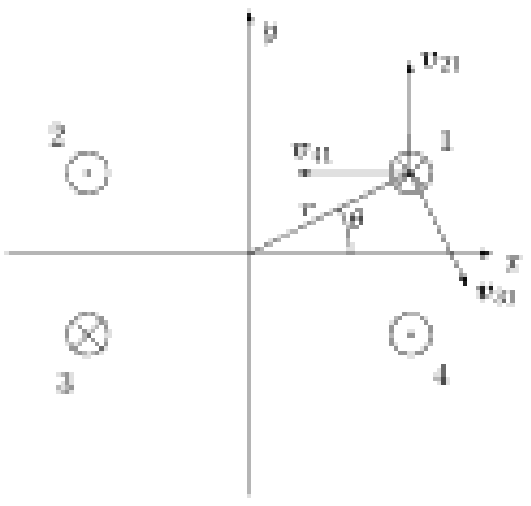}
    (a)
   \end{center}
  \end{minipage}
  \begin{minipage}{0.47\linewidth}
   \begin{center}
    \epsfxsize=\linewidth \epsfbox{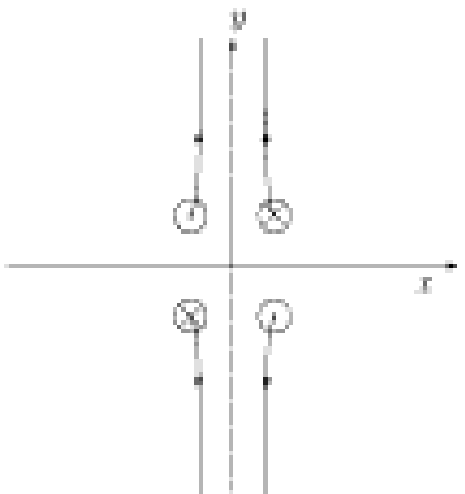}
    (b)
   \end{center}
  \end{minipage}
 \end{center}
 \caption{(a) The initial configuration of Sec.\ref{sound}. The $\odot$
 and $\otimes$ symbols indicate a vortex and an anti-vortex. (b) The
 trajectories of each vortex. the vortex pair above $x$ axis moves upward
 and approaches each other down to a certain distance.}
 \label{fa1}
\end{figure}

This appendix describes the motion of the vortices initially placed as
shown in 
Fig. \ref{fa1} by the vortex filament formulation\cite{lamb}.
The velocity field generated by a
vortex whose singular core is at the origin is given by
\begin{equation}
 v_s=\dfrac{\kappa}{2\pi r},
\end{equation}
where $\kappa$ is the quantized circulation. The motion of equation of
the vortex $1$ is given by
\begin{equation}
 \dot{\mbf{s}}=\mbf{v}_1=\mbf{v}_{21}+\mbf{v}_{31}+\mbf{v}_{41}, 
\end{equation}
where $\mbf{s}=(x,y)$, and $\mbf{v}_{21}$,
$\mbf{v}_{31}$ and $\mbf{v}_{41}$ are the velocity field at the position
of vortex $1$ generated by the vortex $2$, $3$ and $4$, respectively.
Each velocity are given by
\begin{eqnarray}
 \mbf{v}_{21}=\dfrac{\kappa}{2\pi}\left(0,\dfrac1{2x}\right), \\
 \mbf{v}_{31}=\dfrac{\kappa}{2\pi}\dfrac1{2r}\left(\sin\theta,
 -\cos\theta\right), \\
 \mbf{v}_{41}=\dfrac{\kappa}{2\pi}\left(-\dfrac1{2y},0\right).
\end{eqnarray}
Therefore, the dimensionless form of the equations of motion are
\begin{equation}
 \dot{x}=-\dfrac1y+\dfrac{y}{x^2+y^2},\quad
 \dot{y}=\dfrac1x-\dfrac{x}{x^2+y^2},
\end{equation}
where $t\cdot\kappa/2\pi\rightarrow t$.
These equations can be rewritten in term of the polar coordinates:
\begin{equation}
 \dot{r}=-\dfrac2r\cot 2\theta,\qquad \dot{\theta}=\dfrac1{r^2}>0. 
\end{equation}
Therefore
\begin{equation}
 \dfrac{dr}{d\theta}=-2r\cot2\theta.
\end{equation}
For example the vortex $1$ is initially placed at
$(r,\theta)=(r_0,\pi/4)$,
\begin{equation}
 \dfrac{r_0}r=\sin2\theta.
\end{equation}
This equation can be rewritten in term of the Cartesian coordinates:
\begin{equation}
 y^2=\dfrac{\dfrac{r_0^2}4}{x^2-\dfrac{r_0^2}4}
\end{equation}
This solution means that the vortex pair above $x$ axis moves upward
and approaches each other down to a certain distance determined
by the initial configuration as shown in Fig. \ref{fa1} (b),
{\it i.e.} $\Delta L=r_0$.

If $\Delta L$ becomes comparable to the
healing length, the vortex filament formulation becomes not available,
leading to the pair annihilation as described in the text.

%%%%%%%%%%%%%%%%%%%% References %%%%%%%%%%%%%%%%%%%%%%

% figures follow here

% tables follow here

\end{document}